\documentclass[twocolumn]{aastex631}

\usepackage{CJKutf8, color}

\DeclareRobustCommand{\hlsout}{\bgroup\markoverwith{\textcolor{red}{\rule[.5ex]{2pt}{0.4pt}}}\ULon}

\DeclareRobustCommand{\hlsout}{\bgroup\markoverwith{\textcolor{magenta}{\rule[.5ex]{2pt}{0.4pt}}}\ULon}

\usepackage{graphicx}

\usepackage{multirow}
\begin{document}

\title{Systematic study of the composition of Type I X-ray burst ashes: \\ Neutron star structure v.s. Reaction rate uncertainties}

\correspondingauthor{Helei Liu}
\email{heleiliu@xju.edu.cn}
\email{dohi@riken.jp}
\email{r.x.xu@pku.edu.cn}

\author{Guoqing Zhen}
\affiliation{School of Physical Science and Technology, Xinjiang University, Urumqi 830046, China}

\author{Helei Liu}
\affiliation{School of Physical Science and Technology, Xinjiang University, Urumqi 830046, China}

\author{Akira Dohi} 
\affiliation{Astrophysical Big-Bang Laboratory (ABBL), CPR, RIKEN, Wako, Saitama 351-0198, Japan}
\affiliation{Interdisciplinary Theoretical and Mathematical Sciences Program (iTHEMS), RIKEN, Wako, Saitama 351-0198, Japan}

\author{Guoliang L\"{u}}
\affiliation{Xinjiang Astronomical Observatory, Chinese Academy of Science, 150 Science 1-Street, Urumqi 830011, China}
\affiliation{School of Physical Science and Technology, Xinjiang University, Urumqi 830046, China}

\author{Nobuya Nishimura}
\affiliation{Center for Nuclear Study (CNS), The University of Tokyo, Bunkyo-ku, Tokyo 113-0033, Japan}
\affiliation{Astrophysical Big-Bang Laboratory (ABBL), CPR, RIKEN, Wako, Saitama 351-0198, Japan}
\affiliation{National Astronomical Observatory of Japan (NAOJ), Osawa, Mitaka 181-8588, Japan}

\author{Chunhua Zhu}
\affiliation{School of Physical Science and Technology, Xinjiang University, Urumqi 830046, China}

\author{Renxin Xu}
\affiliation{Department of Astronomy, Peking University, Beijing 100871, China}
\affiliation{Kavli Institute for Astronomy and Astrophysics, Peking University, Beijing 100871, China}


\begin{abstract}

In this study, we calculate for the first time the impacts of neutron star(NS) structure on the type I X-ray burst ashes using the \texttt{MESA} code. We find an increased mass fraction of the heavier elements with increasing surface gravity (increase mass or decrease radius), resulting in a higher average mass number ($A_{\rm ash}$) of burst ashes (except for higher mass NS due to the competition between the envelope temperature and the recurrence time). The burst strength ($\alpha$)  increases as surface gravity increases, which indicates the positive correlation between $A_{\rm ash}$ and $\alpha$ with changes in surface gravity. If the $\alpha$ value is higher, heavier $p$-nuclei should be produced by the type I X-ray burst nucleosynthesis. Besides, the effects of various burst input parameters, e.g. base heating ($Q_{\rm b}$), metallicity ($Z$) and some new reaction rates are calculated for comparison.
We find that the heavier nuclei synthesis is inversely correlated to the base heating/metallicity, the smaller the base heating/metallicity, the greater the mass fraction of the heavier elements. The $\alpha$ value decreases as $Q_{\rm b}$ or $Z$ decreases,  which also indicates the positive correlation between $A_{\rm ash}$ and $\alpha$ with variation in $Q_{\rm b}$ or $Z$. The new reaction rates from the $(p,\gamma)$ reactions on  $^{17}\rm{F}$, $^{19}\rm{F}$, $^{26}\rm{P}$, $^{56}\rm{Cu}$, $^{65}\rm{As}$, and $(\alpha,p)$ reaction on $^{22}\rm{Mg}$ have only minimal effects on burst ashes. In hydrogen-rich X-ray binary systems, nuclei heavier than $^{64}\rm{Ge}$ are fertile produced with larger NS mass, smaller NS radius, smaller base heating and smaller metallicity. 

\end{abstract}


\section{Introduction} \label{sec:intro}

Type I X-ray bursts are rapidly brightening phenomena caused by the surface unstable thermonuclear burning of neutron stars(NSs) in low-mass X-ray binaries (LMXBs) \citep{1993SSRv...62..223L,1998ASIC..515..419B,2000A&A...357L..21C, 2002ApJ...566.1045S, 2006NuPhA.777..601S, 2011A&A...525A.111I, 2021ASSL..461..209G}. 
Based on the shape of the light curves, type I X-ray bursts can be classified as standard bursts\footnote{In this paper, we study the standard bursts (i.e., mixed H/He X-ray burst)}, medium/intermediate bursts, and superbursts, where the fuel is mixed hydrogen/helium (H/He), pure helium, and carbon, respectively~(see \cite{2023MNRAS.521.3608A} for the latest observations). In LMXBs, NSs accrete the hydrogen/helium-rich material from their companions via Roche lobe overflow, which heats the NS surface.

In particular, the standard X-ray burst (XRB) with hydrogen-rich (typically $\sim$70\%), which occurs in the case of high accretion rate ($\sim10^{-9}\,\rm{M_{\odot}/yr}$)~\citep{1981ApJ...247..267F,1998ASIC..515..419B}, is often paid attention to the powerful site of the heavy proton-rich nuclei. The nucleosynthesis starts from the 3$\alpha$ reaction around the temperature of 0.2 GK, and hot (bi-)CNO cycles run. After CNO breakout around 0.5 GK, the heavy proton-rich nuclei are synthesized due to $\alpha p$-process and the rapid proton capture process ($rp$-process)~\citep{1981ApJS...45..389W,1998PhR...294..167S,1999ApJ...524.1014S,1999A&A...342..464K}, which terminates at the mass number $A=107$ due to the SnSbTe cycle~\citep{2001PhRvL..86.3471S}~(see \cite{2018JPhG...45i3001M} for a review). For example, \cite{1999ApJ...524.1014S} studied the X-ray burst ashes from the stable nuclear-burning shell with large reaction network including 631 nuclei, fixing $M=1.4\,M_{\odot}$ and $R=10\,{\rm km}$, and showed that the $rp$-process could be the origin of light proton-rich nuclei enough to produce the observed solar abundances, such as $^{84}\rm{Sr}$, $^{92,94}\rm{Mo}$, $^{96,98}\rm{Ru}$, $^{102}\rm{Pd}$, and $^{106}\rm{Cd}$. Thus, the possibility to synthesize proton-rich heavy nuclei beyond $^{56}\rm{Fe}$ has been well-established, at least with the one-zone model, which, however, considers the only burning shell near the surface of accreting NSs.

Since the turn of the century, the properties of XRBs have been studied in the multi-zone model with the modern nuclear reaction network, which at least covers the accreted layer of NSs~\citep{2004ApJS..151...75W,2008ApJS..174..261F} (see also \cite{2004ApJ...603..242K}). They confirm a strong waiting point around $A=64$ through numerical simulation, which stagnates the \textit{rp}-process nucleosynthesis in the multi-zone framework (see also \cite{2008ApJS..178..110P} for the detailed analysis).
This implies that the environment in accreting NSs, which is difficult for the one-zone model to treat, must be important for describing X-ray burst nucleosynthesis.

A number of studies of X-ray bursts have been done in various multi-zone models for the realistic description of light curves~(e.g., \cite{2004ApJS..151...75W,2007ApJ...671L.141H,2016ApJ...830...55C,2018ApJ...860..147M,2021PhRvL.127q2701H}). Thanks to them, it is well understood that the uncertainties of the nuclear reaction rates, the mass accretion rate, and the composition of the accreted matter have been shown to significantly affect the X-ray bursts.
Recently, the microphysics inside accreting NSs has been shown to be another important factor in describing light curves, such as the uncertainties of the nuclear equation of state (EOS), mass, and neutrino cooling process~\citep{2021ApJ...923...64D,2022ApJ...937..124D,2024ApJ...960...14D}, the effect of opacity~\citep{2024arXiv241109843N}. Therefore,
one can get information on NS microphysics as well as on the LMXB through the modeling of light curves of X-ray bursters such as the well-known Clocked bursters GS 1826--24~(e.g., see \cite{2020PTEP.2020c3E02D,2020MNRAS.494.4576J}), in addition to various established approaches to probe the NS interior (for recent reviews, see e.g., \cite{2021PrPNP.12003879B,2024LRR....27....3K}).

The origin of light curves should derive from the compositions of burst ashes since they are originally powered by explosive nuclear burning triggered by hydrogen/helium burning. Hence, the properties of burst ashes have been investigated in the multi-zone models. For instance, \cite{2016ApJ...830...55C} investigated the model-parameter dependence on the burst ashes, focusing on the uncertainties of the $(p, \gamma)$,$(\alpha, \gamma)$ and $(\alpha, p)$ reaction rates based on \texttt{Kepler} code. \cite{2019ApJ...872...84M} also investigated properties of burst ashes with wide parameter regions of accretion rate, compositions, artificial heating rate, and uncertainties of several important reaction rates with the use of \texttt{MESA} code. Note that in their treatment to consider the uncertainty of reaction rates, they vary by a constant value independent of temperature. Hence, it is still necessary to investigate burst ashes with various kinds of modified reaction rates (but see also \cite{2022ApJ...929...72L,2022ApJ...929...73L} for recent progress). Furthermore, they fix the NS mass and radius.
Since the NS microphysics determining the NS mass and radius definitely affects the light curves, as we mentioned above, they must also affect the properties of burst ashes, such as how heavy proton-rich nuclei are synthesized. In fact, 
the burst parameter $\alpha$, which is defined as the ratio between accreting energy to burst energy, tends to be higher if the EOS is softer or the mass is higher~\citep{2021ApJ...923...64D,2023ApJ...950..110Z}. This means that more fuel is available for \textit{rp}-process nucleosynthesis in more compacted NSs. In this work, we investigate the impacts of NS structure and base heating on the properties of X-ray burst ashes. The impacts on light curves have already been reported in the same formulation with the MESA code by our previous work~\citep{2023ApJ...950..110Z}.

The structure of this paper is as follows. Section~\ref{sec:meth} discusses the production of model burst ashes, including the code details and microphysics, as well as some new reaction rates. In Section~\ref{sec:floats}, we explore the composition of burst ashes in various NS masses, NS radii, base heating, metallicity, and some of the new reaction rates. The average mass number of burst ashes, burst strength, impurity parameter, and mean charge number of burst ashes under various changes are discussed in Section~\ref{sec:dis}. Finally, conclusions from our findings in this work are summarized in Section~\ref{sec:con}.

\section{Method} \label{sec:meth}
To study the composition of type I X-ray burst ashes, we use the one-dimensional stellar evolution code \texttt{MESA}, which covers the NS envelope for computation. 
The details of the code and the physical inputs in this work are described as below (see also \cite{2023ApJ...950..110Z}).

\subsection{Code details and microphysics}
The open source stellar evolution code \texttt{MESA} (version 9793) was adopted in this work~\citep{Paxton2011ApJS..192....3P, 2013ApJS..208....4P, Paxton2015ApJS..220...15P, Paxton2018ApJS..234...34P}. To simulate hydrogen-rich X-ray burst, the initial model \texttt{ns\_h} was used, which was taken from \texttt{test\_suite} in folder \texttt{star}.
The same as \cite{2018ApJ...860..147M}, we establish the inner boundary conditions for the NS with mass of $1.4\,M_{\odot}$ and radius of $11.2\,\rm km$, and the envelope thickness of $0.01\,\rm{km}$ with the initial composition of 70\% hydrogen, 28\% helium and 2\% metals, where the command \textit{accretion\_zfracs=3} is set to correspond to the distribution of metals from \cite{1998SSRv...85..161G}. To simulate a typical H/He mixed burst with different metallicity, we set the hydrogen mass fraction of accretion as $X=0.7$ and the helium mass fraction as $Y=0.25, 0.28, 0.29, 0.295$, corresponding the metallicity as $Z=0.05, 0.02, 0.01, 0.005$ from the definition of $X+Y+Z=1$, respectively. 
MESA code simulates the accreting layers above NS solid crust (see Figure 1 in \cite{2023ApJ...950..110Z}), where base heating parameter $Q_{\rm b}$ is adopted at the inner boundary to mimic the energy transfer from the NS interior. In the MESA, this is achieved by fixing the luminosity at the bottom of the envelope layer ($L_{\rm base}=\dot{M}Q_{\rm b}$), so the base luminosity depends on the $Q_{\rm b}$ and $\dot{M}$ of the model. The change of mass, radius and base luminosity by using the commands ``\textit{relax\underline{~}M\underline{~}center}", ``\textit{relax\underline{~}R\underline{~}center}" and ``\textit{relax\underline{~}L\underline{~}center}", respectively. GR effects were taken into account by using a post-Newtonian modification to the local gravity, where the \texttt{MESA} setting ``\textit{use\_GR\_factors = .true.}" was chosen. It's worth mentioning that the data output by MESA is not redshifted to infinity, the light curves and times are local at the surface of NS in the following calculations. Adaptive time and spatial resolution were employed according to the MESA controls \textit{varcontrol\underline{~}target=1d-3} and \textit{mesh\underline{~}delta\underline{~}coeff=0.5{-}1.5}. 
We used the ``\textit{rp.net}" nuclear reaction network consisting of 304 nuclides, and the nuclear reaction rates from the REACLIB V2.2 library. 
The nuclear flow to the Sn-Sb-Te cycle and the production of heavy element ashes were achieved by setting ``\textit{max\underline{~}abar\underline{~}for\underline{~}burning = 120}". 
MESA also provides users with the convenience of modifying the nuclear reaction rate with the set of temperature in units of
$10^{8}\,{\rm K}$ and the rate $N_{A}<\sigma\nu>$ in units of $\rm{cm^{3}\,g^{-1}\,s^{-1}}$.
The details of the parameters setting can be found in \cite{2013ApJS..208....4P}.

\subsection{New reaction rates} \label{sec:rate}
The composition of the burst ashes is sensitive to the reaction rates~\citep{1981ApJS...45..389W, 2004ApJ...608L..61F, 2016ApJ...830...55C, 2021PhRvL.127q2701H, 2022ApJ...929...72L, 2022ApJ...929...73L, 2023ApJ...950..133H, 2024MNRAS.529.3103S}. However, most of the reaction rates occurring in X-ray bursts, in particular, relevant to \textit{rp}-process, have not been directly measured experimentally (but see \cite{2023NatPh..19.1091Z} for recent direct measurement). We therefore focus here on the state-of-the-art nuclear reaction rates determined in recent years to predict the composition of the burst ashes.

The nuclear burning process mainly involved the 3$\rm \alpha$, hot--CNO cycle, $\rm \alpha p$ process and $rp$-process. Therefore, the breakout reactions from the hot-CNO cycle, the $(\alpha,p)$ and $(p,\gamma)$ reactions are of special importance to the X-ray burst nucleosynthesis. \cite{2017PhRvC..96d5812K} obtained the ${}^{\rm 17}{\rm F}(p,\gamma){}^{\rm 18}{\rm Ne}$ reaction rate for the first time from experimental data on $^{18}\rm{Ne}$, which has great influences on the temperature conditions and time scale for breakout from the hot-CNO cycles.
\cite{2021PhRvL.127q2701H} obtained the new ${}^{\rm 22}{\rm Mg}(\alpha,p){}^{\rm 25}{\rm Al}$ reaction rate by performing the first elastic scattering measurement of $^{25}\rm{Al}+p$. However, they only discussed the impact of the new rate on the only light curves with the 1D multizone hydrodynamic \texttt{KEPLER} code, not burst ashes yet. 
\cite{2022ApJ...929...72L,2022ApJ...929...73L} re-assessed the ${}^{\rm 65}{\rm As}(p,\gamma){}^{\rm 66}{\rm Se}$, ${}^{\rm 57}{\rm Cu}(p,\gamma){}^{\rm 58}{\rm Zn}$ reaction rates based on the experimental data and discussed the effects of the new rates on the light curves and the final products using the \texttt{KEPLER} code. 
\cite{2022Natur.610..656Z} measured the ${}^{\rm 19}{\rm F}(p,\gamma){}^{\rm 20}{\rm Ne}$ reaction rate directly from the experimental results obtained in the China JinPing Underground Laboratory (JULA), where they found that the new rate was up to a factor of 7.4 larger than the previous recommended rate. Thus, it is worth investigating the impact of this breakout reaction rate from the hot-CNO cycle on the burst ashes.
\cite{2023ApJ...950..133H} reevaluated ${}^{\rm 26}{\rm P}(p,\gamma){}^{\rm 27}{\rm S}$ reaction rate with use of the experimentally constrained $^{27}\rm S$ mass. They calculated the new forward and reverse rates for $^{26}\rm{P}(p,\gamma)^{27}\rm{S}$ on the distribution of burst ashes by using the one-zone postprocessing nucleosynthesis code \texttt{ppn}, and found that the new rates had no discernible impact on the yield of burst ashes. Based on these new determined reaction rates, we will unified explore the impacts of these new rates on the burst ashes with use of the \texttt{MESA} code.

To date, it has not been directly explored how NS structure uncertainties influence burst ashes. We explore this in the present work, building a series of models based on our previous work~\citep{2023ApJ...950..110Z}. As the impacts of mass accretion rate and hydrogen mass fraction on the composition of the burst ashes have been discussed elsewhere\footnote{For mixed H/He burst, increasing mass accretion rate, increasing hydrogen mass fraction all increase the mass fraction of the heavier elements($A>60$).}~\citep{1999ApJ...524.1014S, 2019ApJ...872...84M}, we explore the impacts of other burst input parameters(e.g. base heating, metallicity and new reaction rates) on type I X-ray burst ashes in this work, focusing on the changes of average mass number and the impurity parameter of burst ashes in light of the varying parameters. The details of the inputs and outputs of our models can be found in Table~\ref{tab:inp} in Appendix~\ref{sec:app}.

\section{Composition of the type I X-ray burst ashes } \label{sec:floats}
\begin{figure}
\centering
	\includegraphics[width=\columnwidth]{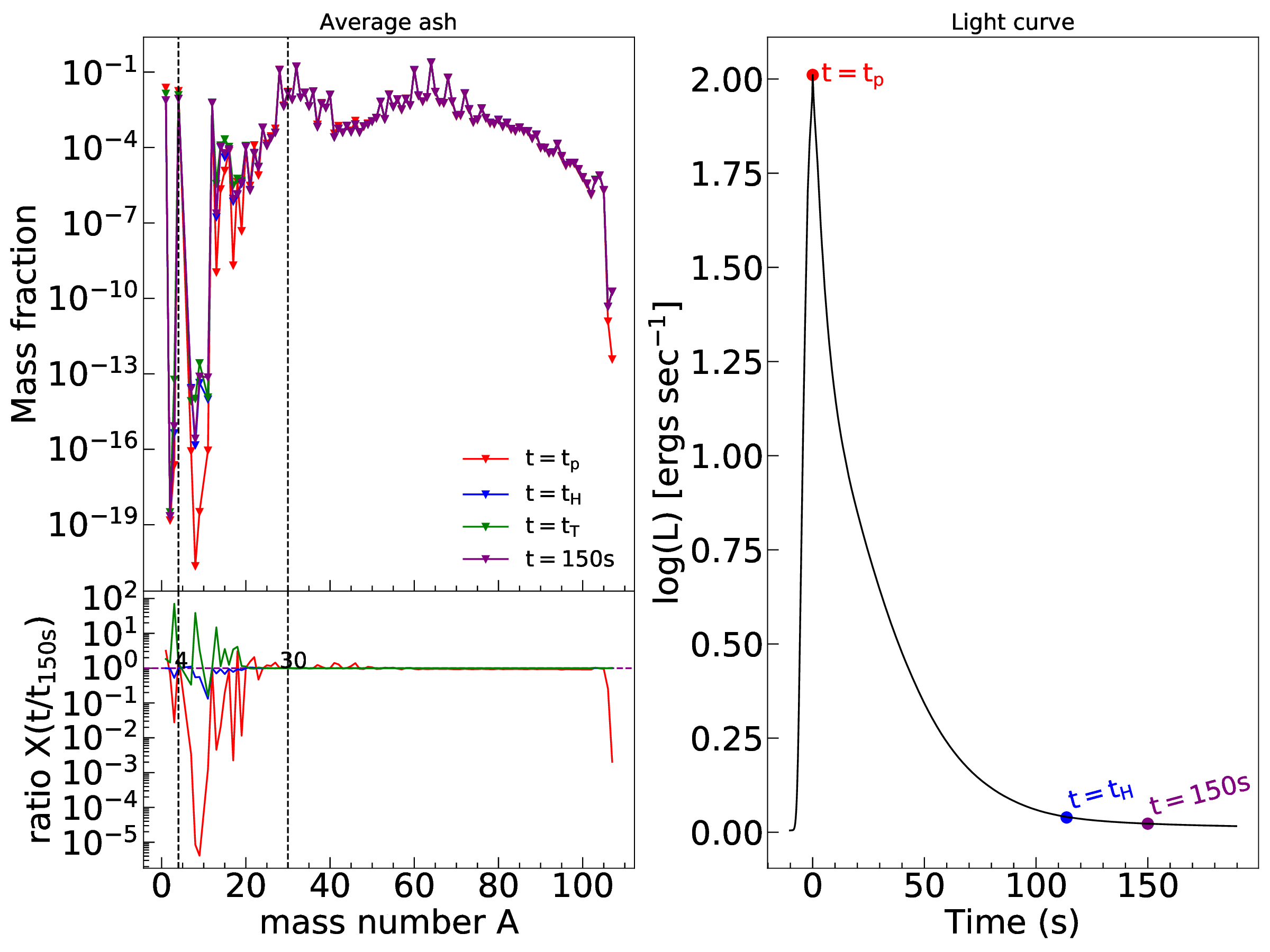}
    \caption{The burst ashes mass fraction distribution as a function of mass number for X-ray burst at $t=t_{\rm p}$ (red),  $t=t_{\rm H}$(blue), $t=t_{\rm T}$ (green) and $t=
    150\,\rm s$ (purple). The lower panel displays the base 10 logarithm of the mass fraction ratio of the $t=t_{\rm p}$ to the $t=t_{\rm H}$ variation (red), the $t=t_{\rm T}$ to the $t=t_{\rm H}$ variation (green) and $t=150\,\rm s$ to the $t=t_{\rm H}$ variation (green). The right panel shows the average light curve, where $t=t_{\rm p}$ is marked by a red dot, $t=t_{\rm H}$ is marked by a blue dot, $t=150\,\rm s$ is marked by a purple dot, at $t_{\rm T}\simeq50$ minutes, as it is beyond the Time-axis, which is not marked. The model input parameters are set as $M=1.4\,M_{\odot}$, $R=11.2\,\rm km$, $X=0.7,Z=0.02$, $Q_{\rm b}=0.1\,\rm{MeV/u}$, $\dot{M}=2\times10^{-9}\,\rm M_{\odot}/yr$.}
    \label{fig:1}
\end{figure}

During a series of X-ray bursts, a large amount of heavy elements ashes are accumulated on the surface of NS. The ashes of earlier bursts will affect the occurrence of subsequent bursts due to the compositional inertia~\citep{1980ApJ...241..358T,2004ApJS..151...75W,2010ApJS..189..204J}, we therefore choose the last burst to calculate the composition of the final products.
Since the ashes produced by the bursts have different proportions of elements at different times, we compare the ashes composition at different times (burst peak time $t=t_{\rm p}$, minimum hydrogen abundance time $t=t_{\rm H}$, minimum effective temperature time $t=t_{\rm T}$ and $t=150\,\rm s$) in Figure~\ref{fig:1}, where the model input parameters are set as $M=1.4\,M_{\odot}$, $R=11.2\,\rm km$, $X=0.7,Y=0.28, Z=0.02$, $Q_{\rm b}=0.1\,\rm{MeV/u}$, $\dot{M}=2\times10^{-9}\,\rm M_{\odot}/yr$ (model 3 in Table~\ref{tab:inp} in Appendix).

The left panel of Figure~\ref{fig:1} shows the mass fraction distribution as a function of mass number at different times, the lower panel displays the variation of the mass fraction ratio, where we choose the baseline as $t=t_{\rm H}$.
The right panel of Figure~\ref{fig:1} shows the average light curve, where $t=t_{\rm p}$ (red dot), $t=t_{\rm H}$ (blue dot) and $t=150\,\rm s$ (purple dot) are marked. Because $t_{\rm T}$ is about $\sim50$ minutes after the burst peak, it can not be seen in the figure.
We find that the proportion of different elements at different times is almost the same between mass number $A\sim 30-100$. The difference of the mass fraction for the lighter elements is mainly due to the accretion.

For convenience, we define the time $t=150\,\rm s$ as the end of the $rp$-process, the composition of the burst ashes is calculated at this moment. The mass fraction of the burst ashes is averaged over the whole multiple radial zones of the envelope, where the accreted material and the ashes from the previous bursts are included.

\begin{figure}
\centering
	\includegraphics[width=\columnwidth]{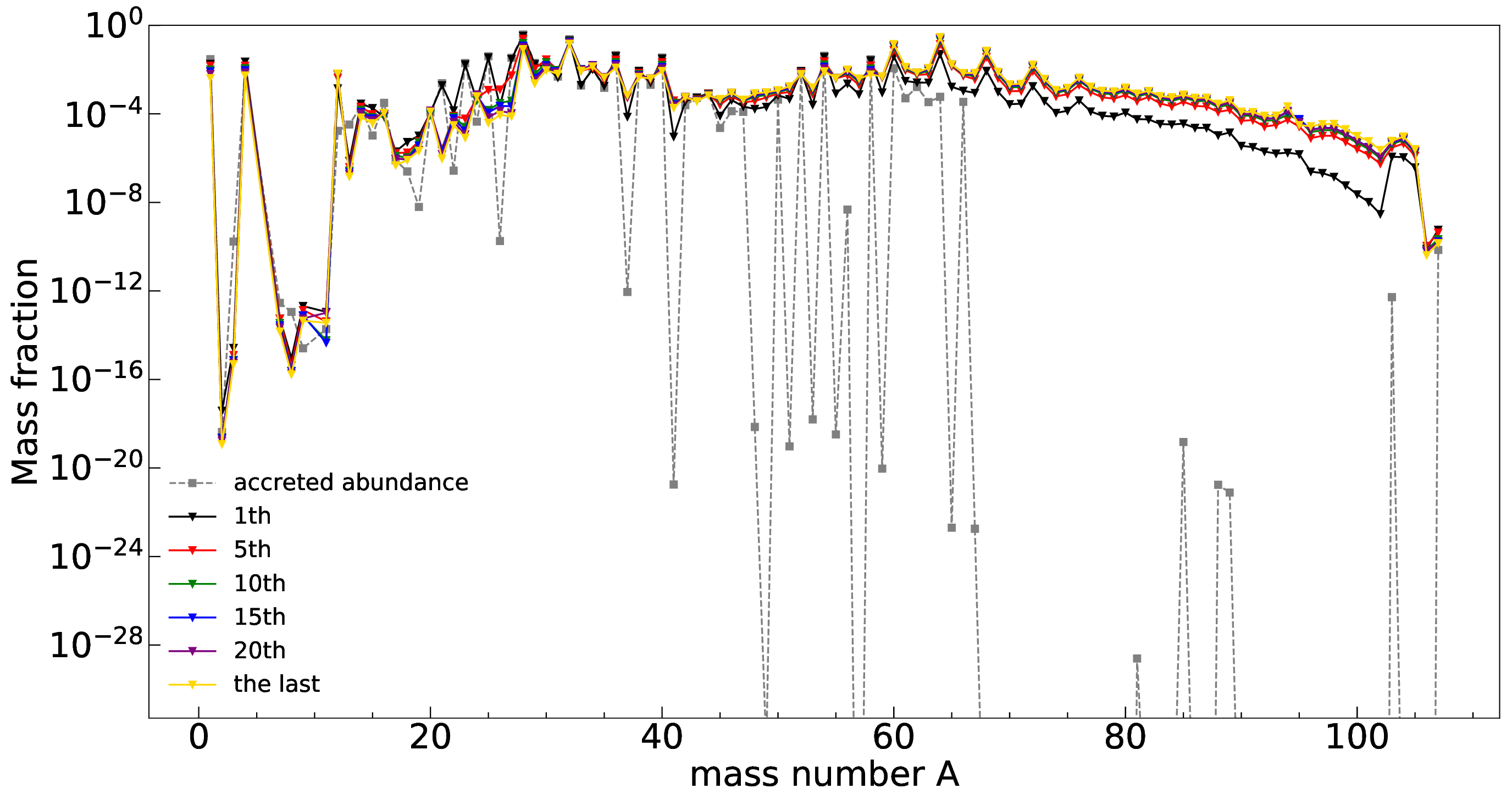}
    \caption{The mass fraction distribution of the burst ashes as a function of mass number for the model calculation with $M=1.4\,M_{\odot}$, $R=11.2\,\rm km$, $X=0.7,Z=0.02$, $Q_{\rm b}=0.1\,\rm{MeV/u}$, $\dot{M}=2\times10^{-9}\,\rm M_{\odot}/yr$. The grey dotted line represents the accreted abundance. The other lines represent the different number of the burst in the burst sequences. Black: the $1\rm th$ burst. Red: the $5\rm th$ burst. Green: the $10\rm th$ burst. Blue: the $15\rm th$ burst. Purple: the $20\rm th$ burst. Gold: the last burst.}
    \label{fig:ash2}
\end{figure}

\subsection{The effects of neutron star mass on the composition of the burst ashes} \label{sec:M}

\begin{figure}
\centering
	\includegraphics[width=\columnwidth]{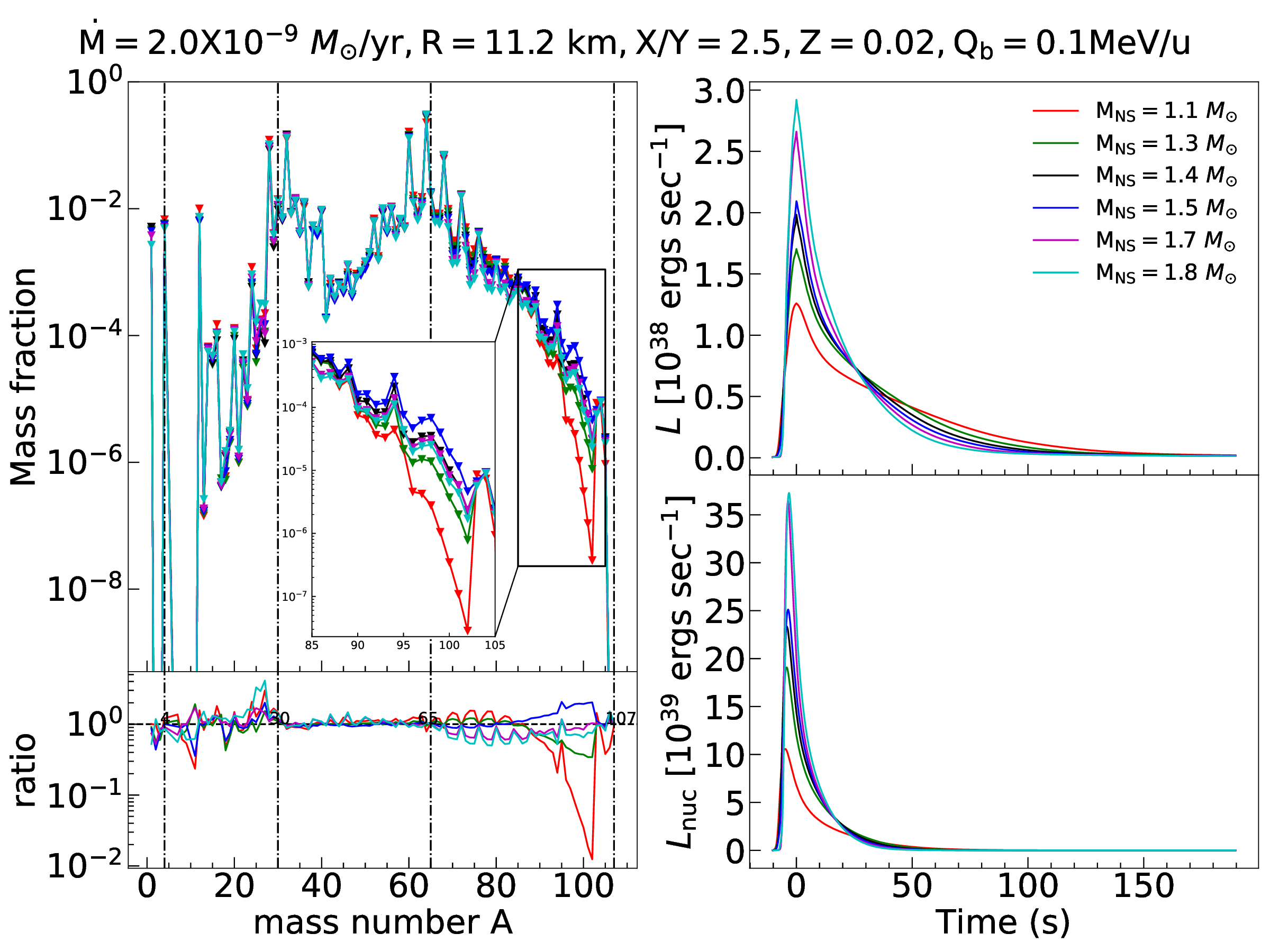}
    \caption{The burst ashes mass fraction distribution as a function of mass number for X-ray burst with changes of NS mass. The lower panel displays the base 10 logarithm of the mass fraction ratio of $M=1.1\,M_{\odot}$ to $M=1.4\,M_{\odot}$ variation (red), $M=1.3\,M_{\odot}$ to $M=1.4\,M_{\odot}$ variation (green), $M=1.5\,M_{\odot}$ to $M=1.4\,M_{\odot}$ variation (blue), $M=1.7\,M_{\odot}$ to $M=1.4\,M_{\odot}$ variation (purple), $M=1.8\,M_{\odot}$ to $M=1.4\,M_{\odot}$ variation (cyan). The right panel shows the total luminosity (up) and the nuclear energy luminosity (down) with different NS masses. The input parameters corresponding models 1-6 in Table~\ref{tab:inp} in the Appendix A.}
    \label{fig:2}
\end{figure}

The mass of NS is an important parameter for studying type I X-ray bursts because mass can significantly affect the surface gravity of NS given as $g_{\rm s}=\frac{GM}{R^2}(1-\frac{2GM}{Rc^2})^{-1/2}$. Therefore, the mass of NS affects the ignition pressure of type I X-ray bursts, which should alter light curves and the composition of the burst ashes. The abundances of burst ashes were extracted after the last burst in the burst sequence, i.e., after $20-30$ bursts. However, we find that the composition of the burst ashes is converged after the $\sim5\rm th$ burst, one can see  Figure \ref{fig:ash2} for detail. The consequences for varying the NS mass are shown in Figure~\ref{fig:2}.
It illustrates that the mass variations that affect the composition of the burst ashes are mainly in the range of mass number $A\sim65-105$. However, limited by our adopted reaction network, the abundance of the p-nuclei such as $^{84}\rm{Sr}$, $^{92,94}\rm{Mo}$, $^{96,98}\rm{Ru}$, $^{102}\rm{Pd}$ can not be deduced.
With increasing the NS mass, the mass fraction of heavier elements increases. The most abundant elements that dominate the composition of the burst ashes (mass fraction $>10^{-2}$) are $A=64, 32, 60, 28, 68, 65, 72, 34, 61, 36, 63$ (for $M=1.4\,M_{\odot}$), from which the odd-A nuclei $A=61,63,65$ are the possible Urca cooling nuclei for the thermal relaxation of quiescent NS transients~\citep{2015PhRvL.115p2501M,2016ApJ...831...13D,2017ApJ...837...73M}. When the mass changes, the most abundant elements are also changed. Details of the results are recorded in Table ~\ref{tab:AX} in Appendix A.
The averaged light curves at different masses can be seen in the right upper panel of Figure \ref{fig:2}.
With the increase of NS mass, the peak luminosity of the light curve increases. Concretely, while the tail luminosity decreases as mass decreases at $t\sim0-30\,\rm s$, the luminosity increases as mass decreases at $30\lesssim t\lesssim 150\,\rm s$ as already shown in \cite{2023ApJ...950..110Z}. 
As there is no monotonicity in the tail part of the light curves with the increase of NS mass, the increase of the heavier elements is inhibited at $M=1.7-1.8\,\rm M_{\odot}$.

\begin{figure}[!htb]
\centering
\includegraphics[width=\columnwidth]{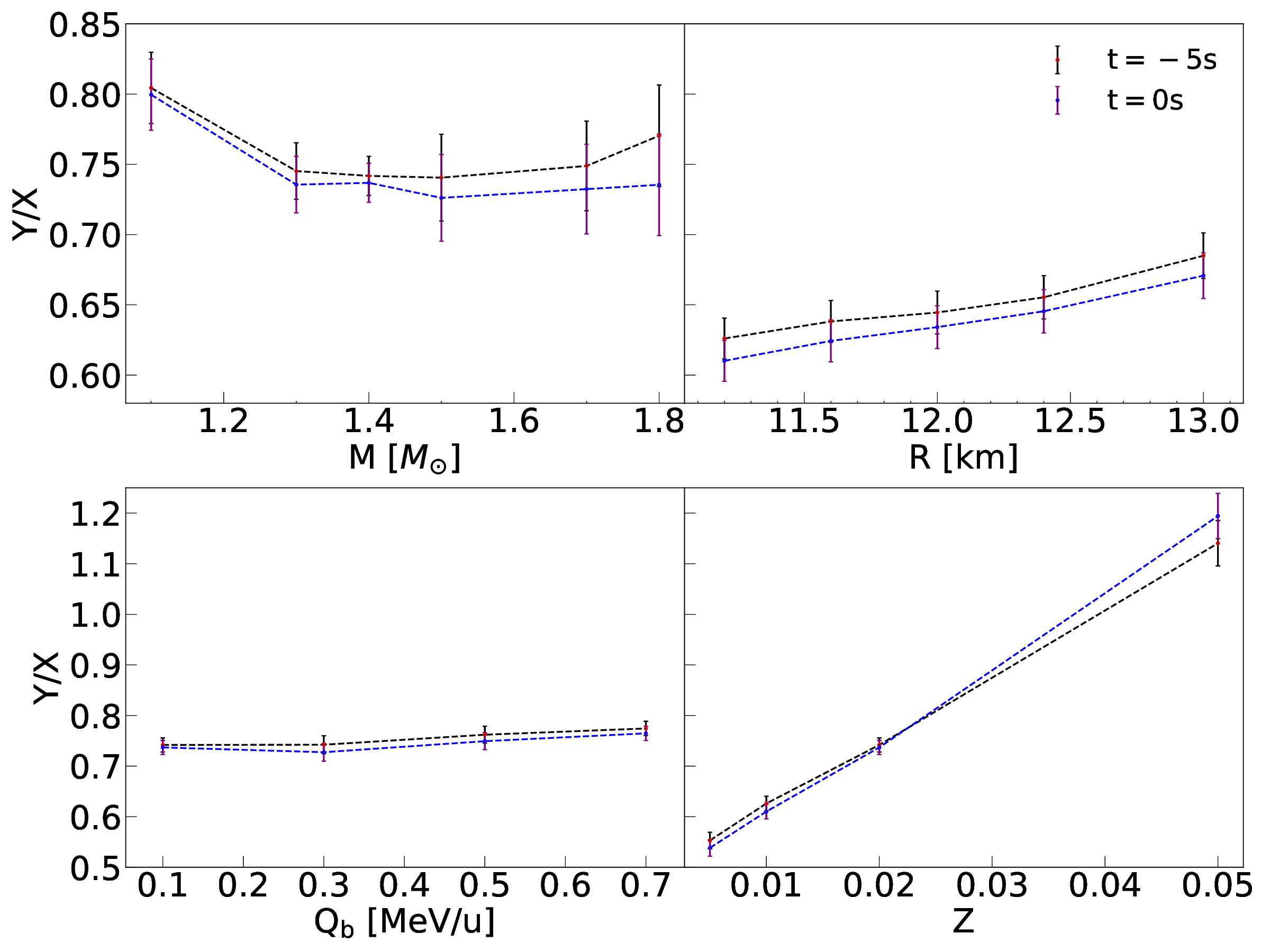} 
\caption{The ratios of $\rm Y/X$ with the variation of NS mass($M$), radius($R$), base heating($Q_{\rm b}$) and metallicity($Z$) at $t=-5\,\rm s$ (five seconds before the burst peak) and $t=0\,\rm s$ (the burst peak).}
\label{fig:YX}
\end{figure}

As shown in the right down panel of Figure \ref{fig:2}, the higher the NS mass is, the higher the peak of nuclear energy luminosity is. This is because the recurrence time increases as NS mass increases(~\citep{2023ApJ...950..110Z}, see also table \ref{tab:inp} in Appendix), which results in more accumulated fuel between bursts and more energetic bursts. The long recurrence time and high temperature will enhance the $rp$-process nucleosynthesis. On the other hand, the ratios of $\rm Y/X$ at the start of the burst is calculated in Figure \ref{fig:YX}, it shows that the value of $\rm Y/X$ decreases as NS mass increases, but for  $M=1.7\,M_{\odot},\, 1.8M_{\odot}$ cases. This is because the temperature of the envelope is higher for the larger mass NS (see also Figure ~\ref{fig:9} for detail), the high temperature (which relatively means lower ignition temperature or higher ignition pressure) results in an earlier arrival at the temperature that He abundance required to trigger the $3\alpha$ reaction for burst ignition. Since the higher neutron star mass leads to higher ignition pressure, less H is required to be burned prior to burst ignition, resulting in a smaller Y/X at burst ignition. On the other hand, the timescale for the accreted hydrogen to be depleted by hot-CNO cycle is $t_{\rm CNO}=9.8\,\rm{hr}(\frac{X}{0.7})(\frac{Z}{0.02})^{-1}$~\citep{2016ApJ...819...46L}. We can get $t_{\rm CNO}=9.8\,\rm hr$ for $X=0.7,Z=0.02$. The recurrence time is in the range of $\sim3.47-3.83\,\rm hr$ with the variation of NS mass (see Table \ref{tab:inp} for detail), which is less than the H depletion time. Although the impact of NS mass on the recurrence time is complicated (see also ~\cite{2021ApJ...923...64D}), this may cause $\rm Y/X$ to decrease as NS mass increases.
From the above analysis, the competition of the temperature of the NS envelope and the recurrence time makes the complex of the changes of $\rm Y/X$ and the nucleosynthesis of the heavier elements as the NS mass increases.


\subsection{The effects of neutron star radius on the composition of the burst ashes}
\begin{figure}[h]
\centering
	\includegraphics[width=\columnwidth]{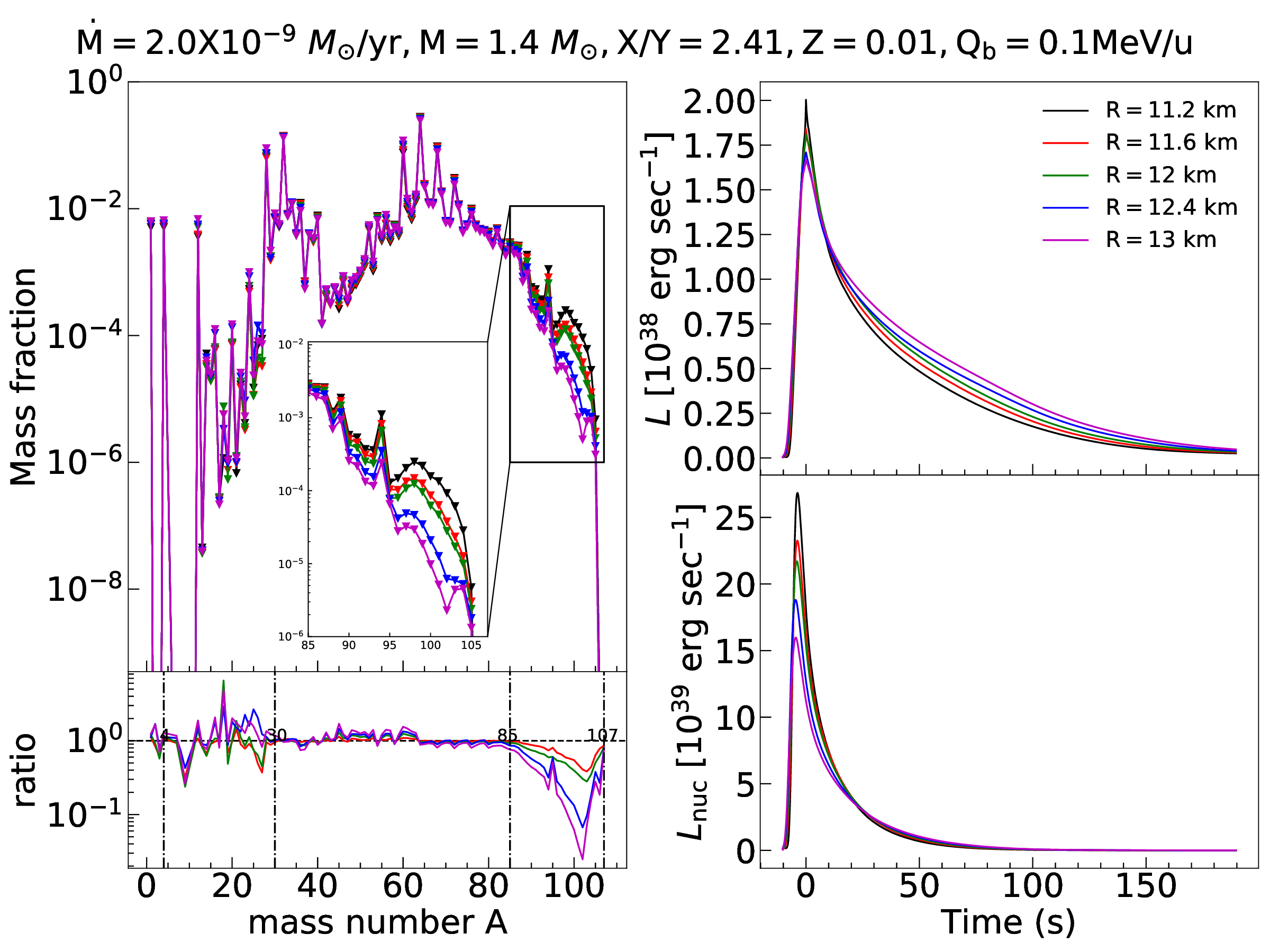}
    \caption{The burst ashes mass fraction distribution as a function of mass number for X-ray burst with changes of NS radius. The lower panel displays the base 10 logarithm of the mass fraction ratio of $R=11.6\,\rm km$ to $R=11.2\,\rm km$ variation (red), $R=12\,\rm km$ to $R=11.2\,\rm km$ variation (green), $R=12.4\,\rm km$ to $R=11.2\,\rm km$ variation (blue), $R=13\,\rm km$ to $R=11.2\,\rm km$ variation (purple). The right panel shows the total luminosity (up) and the nuclear energy luminosity (down) with different NS radii. The input parameters corresponding models 7-11 in Table~\ref{tab:inp} in the Appendix A.}
    \label{fig:3}
\end{figure}

 The abundance of ashes with variations in NS radius is shown in the left panel of Figure \ref{fig:3}. 
As the NS radius increases, the mass fraction of the heavier burst ashes decreases significantly. In the radius of $11.2 -13\,\rm{km}$, the mass fraction of the heavier ashes decreases with the increase of NS radius, the changes are mainly concentrated in the range of $A\sim85-107$.

From the right upper panel of Figure~\ref{fig:3}, we can see that the peak luminosity decreases as the radius increases. The right down panel of Figure~\ref{fig:3} shows that the peak of the nuclear energy luminosity decreases as the radius increases. Increasing radius has the effect of decreasing the envelope temperature especially at high density (see the $T$-$\rho$ diagram in Figure \ref{fig:10} for detail), which inhibits the $rp$-process nucleosynthesis, exhibiting less energetic bursts. From Figure \ref{fig:YX}, we can see that the value of $\rm Y/X$ increases as radius increases,
which means that more hydrogen will be burned prior to burst ignition as radius increases. Hence, the XRBs with larger NS radii are likely to be He-rich bursts. Because less hydrogen is left, those bursts will produce less heavier elements.

In light of the variations of NS mass and radius, we can conclude that the more compact the NS, the more heavier burst ashes are produced. This is physically because of the NS structure's dependence on ignition temperature. For the temperature evolution of bursting NSs, we will mention in Section~\ref{sec:dis}.

\subsection{The effects of base heating on the composition of the burst ashes}
\begin{figure}[h]
\centering
	\includegraphics[width=\columnwidth]{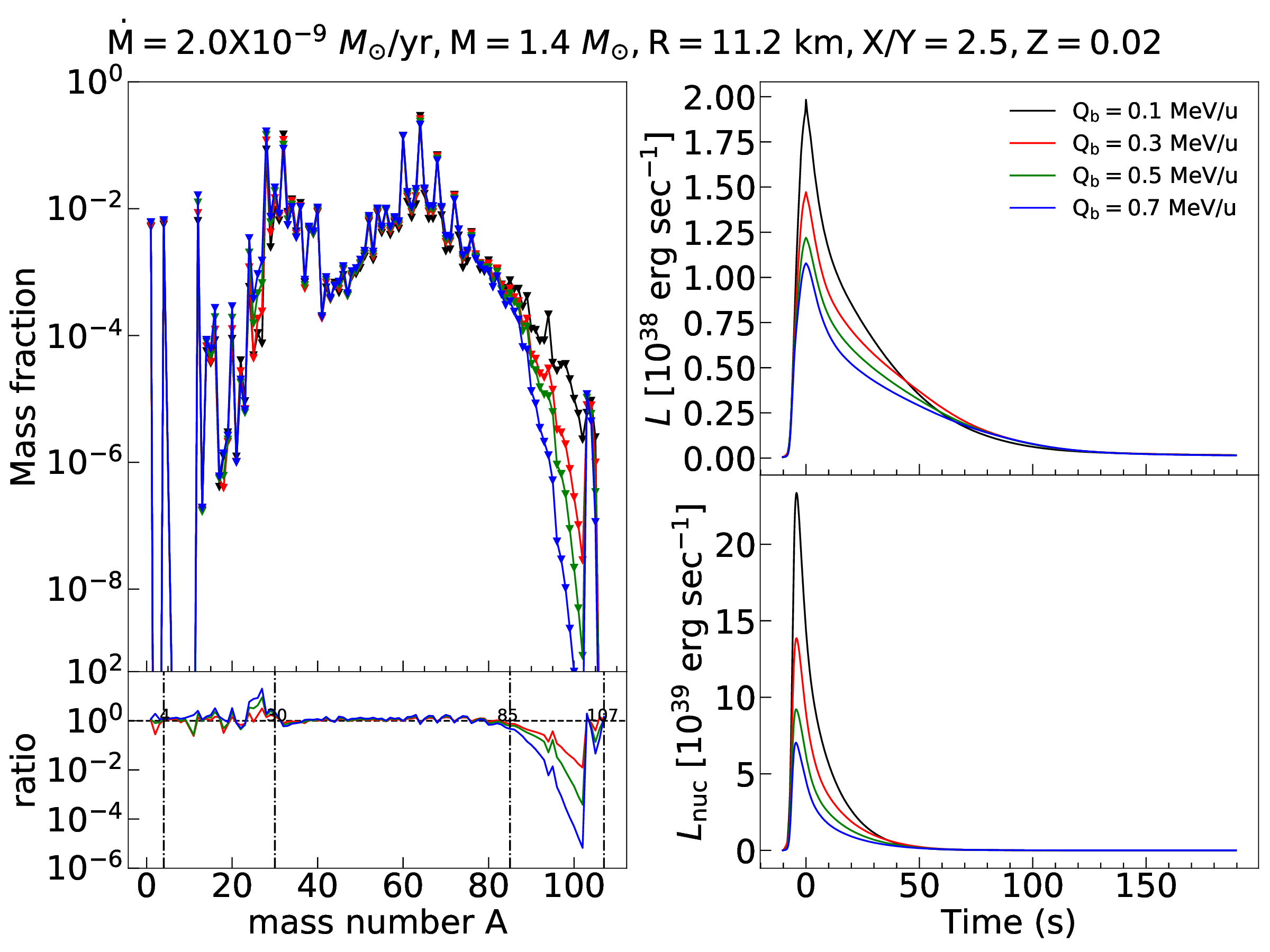}
    \caption{The burst ashes mass fraction distribution as a function of mass number for X-ray burst with changes of base heating $Q_{\rm b}$. The lower panel displays the base 10 logarithm of the mass fraction ratio of $Q_{\rm b}=0.3\,\rm MeV/u$ to $Q_{\rm b}=0.1\,\rm MeV/u$ variation (red), $Q_{\rm b}=0.5\,\rm MeV/u$ to $Q_{\rm b}=0.1\,\rm MeV/u$ variation (green), $Q_{\rm b}=0.7\,\rm MeV/u$ to $Q_{\rm b}=0.1\,\rm MeV/u$ variation (blue). The right panel shows the total luminosity (up) and the nuclear energy luminosity (down) with different base heating. The input parameters corresponding models 3,12-14 in Table~\ref{tab:inp} in the Appendix A.}
    \label{fig:4}
\end{figure}

Base heating parameter $Q_b$ represents the strength of heat flow from the NS crust to the envelope. Following the choice in previous XRB studies~\citep{2018ApJ...860..147M, 2009ApJ...698.1020B, 2017ApJ...842..113K, 2021ApJ...923...64D}, we choose $Q_{\rm b}=0.1,0.3, 0.5, 0.7\,\rm{MeV/u}$. The abundance of ashes with variation in base heating is shown in the left panel of Figure~\ref{fig:4}.
We can see that nuclei heavier than $^{85}\rm Ru$ are more synthesized in smaller-$Q_b$ value, i.e., colder NSs. For $Q_{\rm b}=0.7\,\rm{MeV/u}$, the mass fraction of the final products near mass number $A=102$ is 5 orders of magnitude smaller than that of $Q_{\rm b} = 0.1\,\rm{MeV/u}$.

The corresponding average light curves in the variation of $Q_b$ can be seen in the right upper panel of Figure~\ref{fig:4}. We can see that the peak luminosity decreases as $Q_b$ increases. The tail luminosity also monotonously decreases as $ Q_b$ decreases. The right down panel illustrates that the peak of the nuclear energy luminosity decreases as $Q_b$ decreases, this is because the envelope temperature increases as $Q_{\rm b}$ increases, which leads to the short burning time required to achieve ignition. As less fuel will be accumulated between bursts,  it will show less energetic bursts. Meanwhile, the short recurrence time 
will inhibit the $rp$-process nucleosynthesis and seriously affect the accumulation of heavier ashes. Figure \ref{fig:YX} shows that the value of $\rm Y/X$ increases as $Q_{\rm b}$ increases, which means that more hydrogen will be burned prior to burst ignition as $Q_{\rm b}$ increases. Hence, the XRBs with larger $Q_{\rm b}$ are likely to be He-rich bursts, which produces less heavier elements. From the above analysis, we can understand that the higher peak luminosity and longer tail parts of the light curves are positively correlated to the synthesis of heavier elements.

\subsection{The effects of metallicity on the composition of the burst ashes}

\begin{figure}
\centering
	\includegraphics[width=\columnwidth]{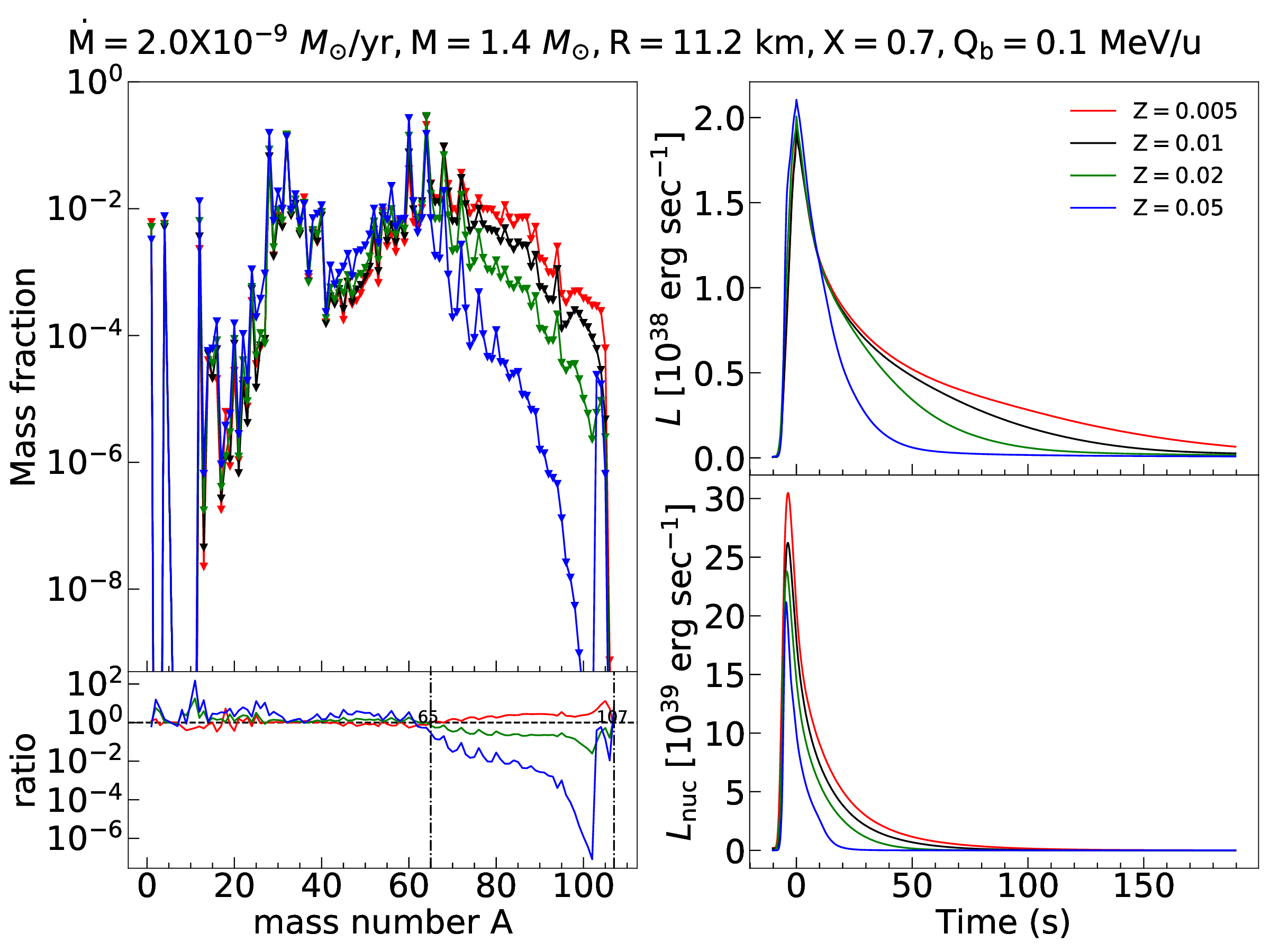}
    \caption{The burst ashes mass fraction distribution as a function of mass number for X-ray burst with variation in metallicity $Z$. The lower panel displays the base 10 logarithms of the mass fraction ratio of $Z=0.005$ to $Z=0.01$ variation (red), $Z=0.02$ to $Z=0.01$ variation (green), $Z=0.05$ to $Z=0.01$ variation (blue). The right panel shows the total luminosity (up) and the nuclear energy luminosity (down) with different metallicity. The input parameters corresponding models 3,7,15,16 in Table~\ref{tab:inp} in the Appendix A.}
    \label{fig:5}
\end{figure}

Since the metallicity provides a catalyst for the hot-CNO cycle, it affects the hydrogen abundance at burst ignition. Higher metallicity leads to hydrogen consumption at the base of the accreted layer long before the burst ignition. Following the choice in previous XRB studies~\citep{2018ApJ...860..147M,2019ApJ...872...84M,2016ApJ...819...46L,2024MNRAS.529.3103S}, we adopted $Z=0.005,0.01, 0.02, 0.05$.
The mass fraction of burst ashes as a function of accretion rate for different metallicities are shown in Figure~\ref{fig:5}. The mass fraction of heavier elements decreases as metallicity increases. This is because there will be less H left to burn during the burst for a higher metallicity case, which inhibits the $rp$-process for heavier elements synthesis. At $Z=0.05$, ashes around A=102 are about six orders of magnitude lower than the case with $Z=0.01$.

The average light curves in the variation of metallicity are shown in the right upper panel of Figure~\ref{fig:5}. We can see that the peak luminosities increase as metallicity increases. However, at the tail part, the luminosity decreases as metallicity increases.
The nuclear energy luminosity is exhibited in the right down panel of Figure~\ref{fig:5}. It shows that the peak luminosity decreases as metallicity increases. Because less hydrogen is left for $rp$-process as metallicity increases, those bursts will be less energetic, exhibiting a lower nuclear energy luminosity. Figure \ref{fig:YX} illustrates that the value of $\rm Y/X$ increases as $Z$ increases, which means more hydrogen will be burned prior to burst ignition as $Z$ increases. Hence, the XRBs with larger $Z$ are likely to be He-rich bursts, resulting in the nucleosynthesis of less heavier elements.
Our analysis illustrates that the longer tail and the higher luminosity during the burst decay are positively correlated to the synthesis of heavier elements.

\subsection{The effects of new reaction rates on the composition of the burst ashes}

We calculate the composition of the burst ashes from the $(p,\gamma)$ reactions on $^{17}\rm{F}$, $^{19}\rm{F}$, $^{26}\rm{P}$, $^{56}\rm{Cu}$, $^{65}\rm{As}$, and $(\alpha,p)$ reaction on $^{22} \rm{Mg}$, which have been declared in Section~\ref{sec:rate}. These new reaction rates are determined from the experiment constraint and model calculations. We focus on the impacts of these new rates on the composition of burst ashes with use of the \texttt{MESA} code.

\begin{figure}[h]
\centering
	\includegraphics[width=\columnwidth]{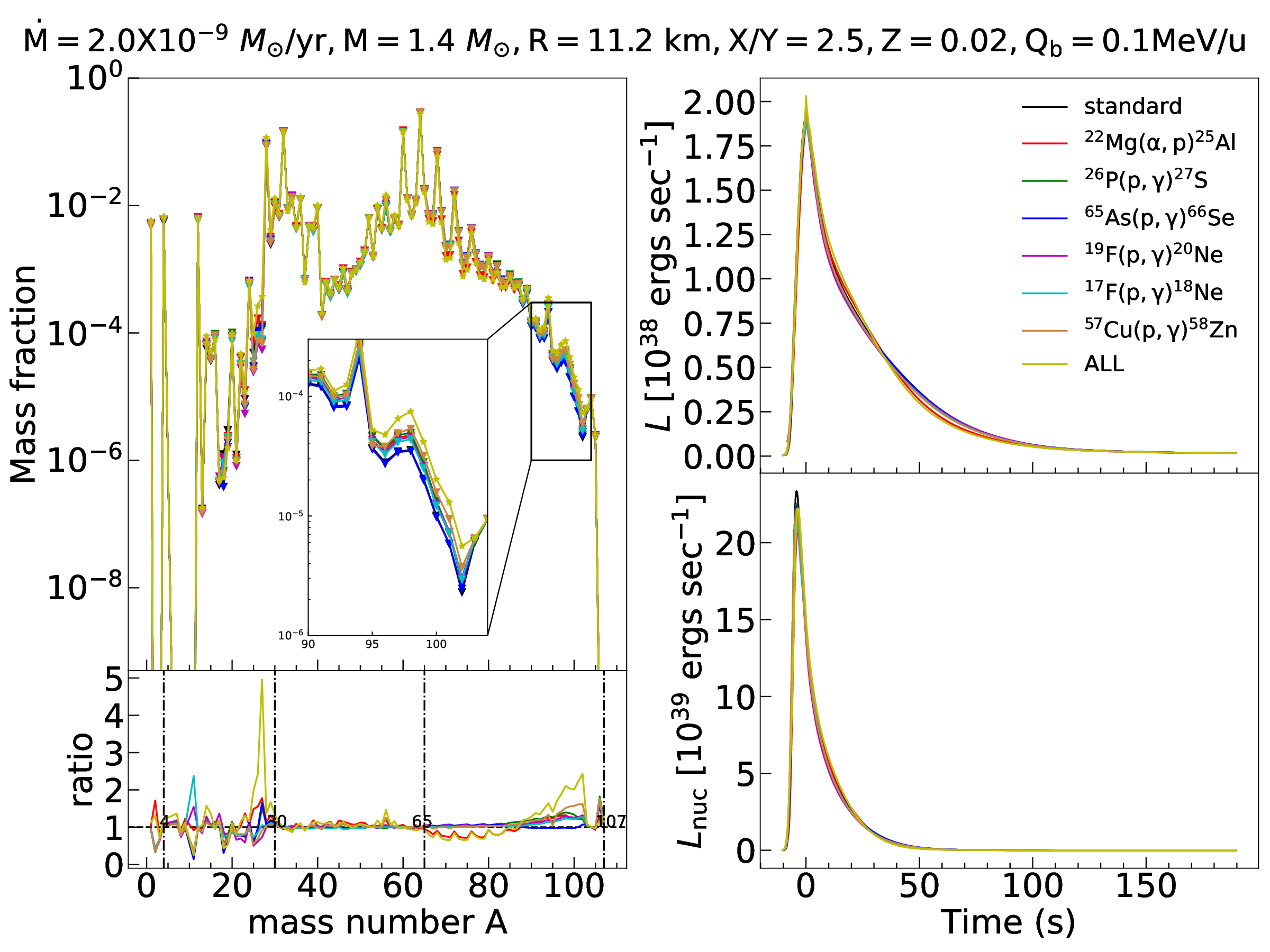}
    \caption{The burst ashes mass fraction distribution as a function of mass number for X-ray burst with changes of some new reaction rates. The lower panel displays the base 10 logarithm of the mass fraction ratio from the new ${}^{\rm 65}{\rm As}(p,\gamma){}^{\rm 66}{\rm Se}$, ${}^{\rm 22}{\rm Mg}(\alpha,p){}^{\rm 25}{\rm Al}$,  ${}^{\rm 17}{\rm F}(p,\gamma){}^{\rm 18}{\rm Ne}$, ${}^{\rm 26}{\rm P}(p,\gamma){}^{\rm 27}{\rm S}$, ${}^{\rm 19}{\rm F}(p,\gamma){}^{\rm 20}{\rm Ne}$,${}^{\rm 57}{\rm Cu}(p,\gamma){}^{\rm 58}{\rm Zn}$ reaction rates to the old reaction rates variations. The right panel shows the total luminosity (up) and the nuclear energy luminosity (down) with the changes of different new reaction rates.}
    \label{fig:6}
\end{figure}

The consequences for varying the reaction rates are illustrated in Figure~\ref{fig:6}. The adoption of the new reaction rates has a tiny impact on the yield of burst ashes compared with the variations in NS mass, radius, base heating, and metallicity.
Among the nuclear reaction rates we changed, the key reactions ${}^{\rm 17}{\rm F}(p,\gamma){}^{\rm 18}{\rm Ne}$, ${}^{\rm 19}{\rm F}(p,\gamma){}^{\rm 20}{\rm Ne}$ that breakout the hot-CNO cycle have the relatively obvious impact on the burst ashes, the maximum mass fraction ratio from the new ${}^{\rm 17}{\rm F}(p,\gamma){}^{\rm 18}{\rm Ne}$, ${}^{\rm 19}{\rm F}(p,\gamma){}^{\rm 20}{\rm Ne}$ reaction rates to the old ones are 1.70 and 2.37, respectively. We obtain that the final abundance ratio of $^{25} \rm{Al}/^{22} \rm{Mg}$ using the new rate is $\mathrm{1.18}$ of that from the old rate, and the maximum mass fraction ratio from the new $^{22} \rm{Mg}(\alpha,p)^{25}\rm{Al}$ to the old one is $\mathrm{1.78}$. The mass fraction of the burst ashes calculated from the new $^{26}\rm{P}(p,\gamma)^{27}\rm{S}$, $^{65}\rm{As}(p,\gamma)^{66}\rm{Se}$, $^{57}\rm{Cu}(p,\gamma)^{58}\rm{Zn}$ reaction rates are consistently with the previous work~\citep{2023A&A...678A.156H, 2022ApJ...929...72L,2022ApJ...929...73L}. 

The average light curves in the variation of the new reaction rate are shown in the right panel of Figure~\ref{fig:6}; we find that the new rates have slight effects on the light curves compared with the effects of other parameters ($M,R,Q_b,Z_{\rm CNO}$), which are consistent with the changes of the mass fraction of burst ashes.

\section{Discussion} \label{sec:dis}

In the process of X-ray bursts, the ashes generated by the burst will be continuously compressed into the crust by the newly accreted matter. Studying the composition of the ashes is conducive to study the composition of the crust, which can provide guidance for the construction of the equation of state of the NS crust and the thermodynamic evolution of the crust~\citep{2018JPhG...45i3001M, 2021MNRAS.507.3860S, 2023MNRAS.523.4643S,2021A&A...645A.102P,2021PhRvL.127q2702W, 2018ApJ...859...62L}.
However, when studying the properties of the crust, some parameters such as average charge, impurity parameter, and average mass number of burst ashes are usually needed.
Here we introduce these parameters to study the crust composition from the X-ray burst ashes.
The average charge number can be expressed as 
\begin{equation} \label{eq:Z}
\langle Z\rangle=\frac{1}{n_{\rm N}}\sum_{i} n_i Z_i , 
\end{equation}
where $Z_i$ and $n_i$ denote charge number and density number of nuclei of the $i$th kind, and $n_{\rm N}=\sum_{i}n_i$ is the total ionic density.
The impurity parameter is used to describe the degree of promiscuity
\begin{equation} \label{eq:Qimp}
Q_{\rm imp}=\frac{1}{n_{\rm N}}\sum_{i} n_i (Z_i-\langle Z\rangle)^{2} . 
\end{equation}
The mean mass number of burst ashes is defined by
\begin{equation} \label{eq:Aash}
A_{\rm ash}=\frac{1}{n_{\rm N}}\sum_{i} n_i A_i , 
\end{equation}
where $A_{\rm i}$ is the mass number of the $i$th nucleus. It should be noticed that the values of $X_{\rm i}$, $Z_{\rm i}$ and $A_{\rm i}$ are averaged from the accreted layer, thus our obtained values of $\langle Z \rangle$, $Q_{\rm imp}$ and $A_{\rm ash}$ are the averaged ones for the envelope.
Besides, the burst strength\footnote{We define $\alpha$ as burst strength. Since for higher $\alpha$ values, $rp$-nuclei is accumulated thanks to more amount of fuel (due to high accretion rate and long recurrence time).
} $\alpha=\frac{z_{g}}{1+z_{g}} \dot{M} c^{2} \frac{\Delta t}{E_{b}}$ is calculated, where $z_{g}$ is the gravitational redshift and $E_b=\int L_b dt$.

\begin{figure}
\centering
	\includegraphics[width=\columnwidth]{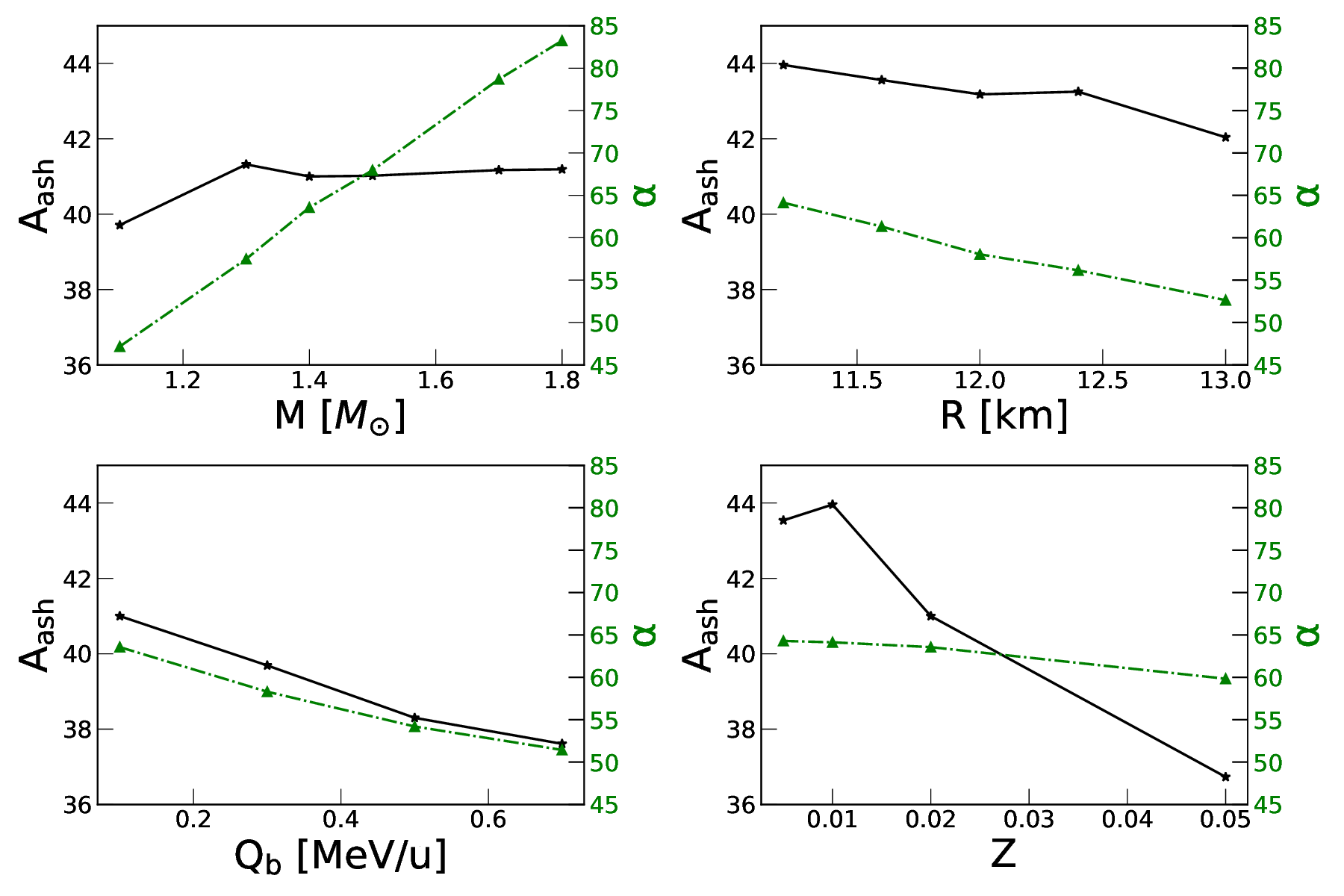} 
    \caption{Values of $A_{\rm ash}$ and $\alpha$ with different M (top left),
R (top right), $Q_b$ (lower right), and Z(lower left). }
    \label{fig:7}
\end{figure}

\begin{figure}
\centering
	\includegraphics[width=\columnwidth]{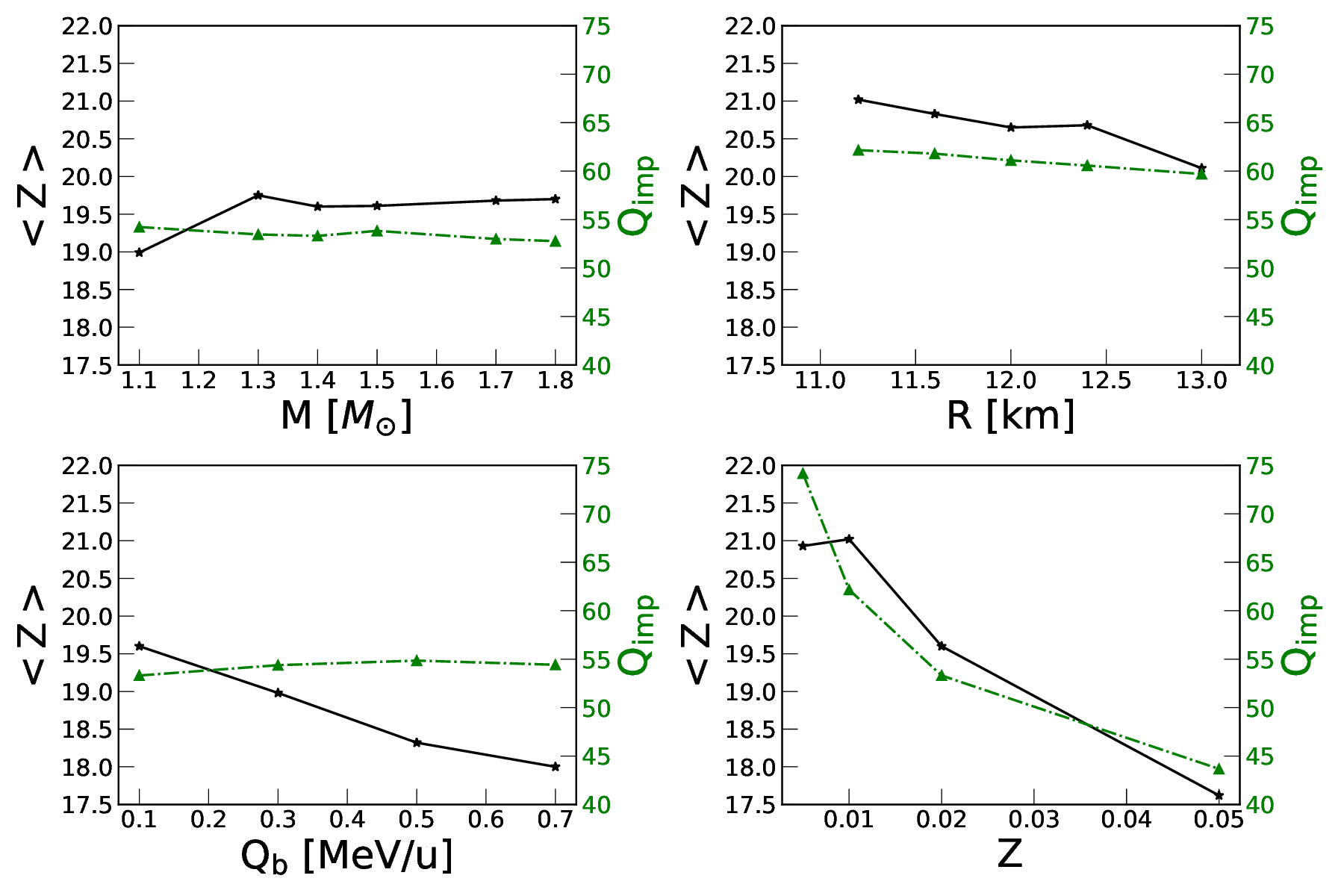} 
    \caption{Values of $<Z>$ and $Q_{\rm imp}$ with different M (top left),
R (top right), $Q_b$ (lower right), and Z(lower left). }
    \label{fig:8}
\end{figure}

The changes of the average mass number of burst ashes ($A_{\rm ash}$), burst strength($\alpha$), average charge number ($\langle Z \rangle$), impurity parameters ($Q_{\rm{imp}}$) with variations in NS mass, radius, base heating and metallicity are explored in the present work. The results can be seen in Figure \ref{fig:7} and Figure \ref{fig:8}.

In Figure \ref{fig:7}, as the mass of NS increases, both the burst strength and the average mass number of the burst ashes increase. As the radius of the NS increases, the burst strength decreases, the average mass number of the burst ashes decreases, which means that the greater the surface gravity, the greater the average mass number and the burst strength.
However, both the mean mass number of the burst ashes and the burst strength decrease as the metallicity or base heating increases. As more heavy elements synthesize at higher surface gravity, lower base heating, and lower metallicity, the changes of $A_{\rm ash}$ are consistent with the changes of the composition of the burst ashes explored in Section~\ref{sec:floats}. We show that the value of $A_{\rm ash}$ is larger for a larger $\alpha$. This implies that heavier nuclei should be synthesized in stronger bursts. On the other hand, from the expression of $\alpha$ value, assuming constant $E_b$ with $M$ and $R$ (but see also Figure 13 in \cite{2021ApJ...923...64D}), the compacted NSs seem to have higher $\alpha$ (see also Figure 5 in \cite{2023ApJ...950..110Z}), which leads to the production of heavier nuclei.

Figure \ref{fig:8} shows the average charge number($\langle Z\rangle$) and the impurity parameter ($Q_{\rm imp}$) of the final products. It is shown that as the mass (radius) of the NS increases (decreases), the average charge number increases, while the impurity parameter have very slightly changes. This means that the more compact the NS is, the more significant the average charge number is, while the variation of the charge distribution is not obviously (as $Q_{\rm imp}$ is almost constant). 
As $Q_{\rm b}$ increases, $\langle Z\rangle$ decreases, $Q_{\rm imp}$ increases. As metallicity increases, both the average charge number $\langle Z \rangle$ and impurity parameter $Q_{\rm imp}$ decrease. Hence, the changes of $\langle Z\rangle$ is consistent with the parameters ($M$, $R$, $Q_{\rm b}$, $Z$) dependence of $A_{\rm ash}$.

The values of $A_{\rm ash}$, $\langle Z\rangle$ and $Q_{\rm imp}$ in the variation of $M$, $R$, $Q_{\rm b}$ and metallicity can be seen in Table~\ref{tab:inp} in the Appendix A for detail. As we see, the effects of NS structure(mass and radius) on these values are not as significant as base heating and metallicity. Especially, the values of $Q_{\rm imp}$ range in $\sim 45-75$ as metallicity varied from 0.005 to 0.05.
For the strong photospheric radius expansion burst, the ashes have been identified from the spectral edges~\citep{2010A&A...520A..81I, 2017ApJ...842..113K, 2021PASJ...73.1405I}. 
It has been suggested that the ejecta of the ashes produced by XRB nucleosynthesis could account for the galactic abundances of some of the light $p$-nuclei that are underproduced in standard astrophysical sites such as the massive star's explosion \citep{1995A&A...298..517R} and Type-Ia supernova \citep{1991ApJ...373L...5H}.

Moreover, in accreting NSs, the thermal and electrical conductivities for the crust will be different under various conditions, leading to a different thermal structure in the NS crust. The quiescent cooling curves observed in several soft X-ray transients provide us with important data on the thermal evolution of the NS outer layers~\citep{2017JApA...38...49W}, thus constraint on the uncertainties about the crust structure of accreting NS. Our calculations provide essential information for the further study of the above issues. 

\begin{figure}
\centering
\includegraphics[width=\columnwidth]{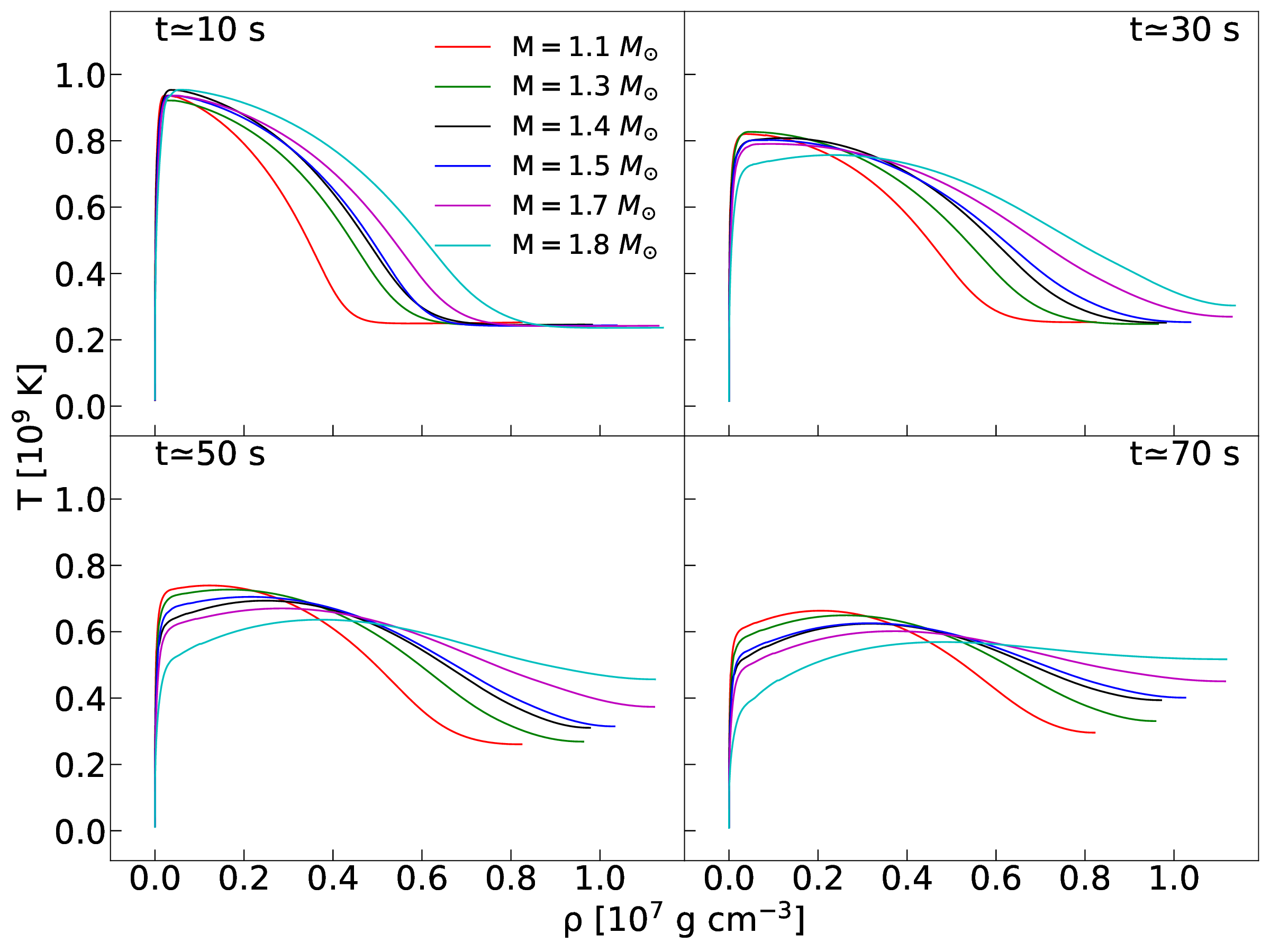} 
\caption{Structure changes in the temperature($T$)-density($\rho$) diagram after the peak of the burst ($t\simeq10,30, 50, \rm and~70\,s$) during a burst cycle. The different curves in a panel show the variation of NS masses.
}
\label{fig:9}
\end{figure}

\begin{figure}
\centering
\includegraphics[width=\columnwidth]{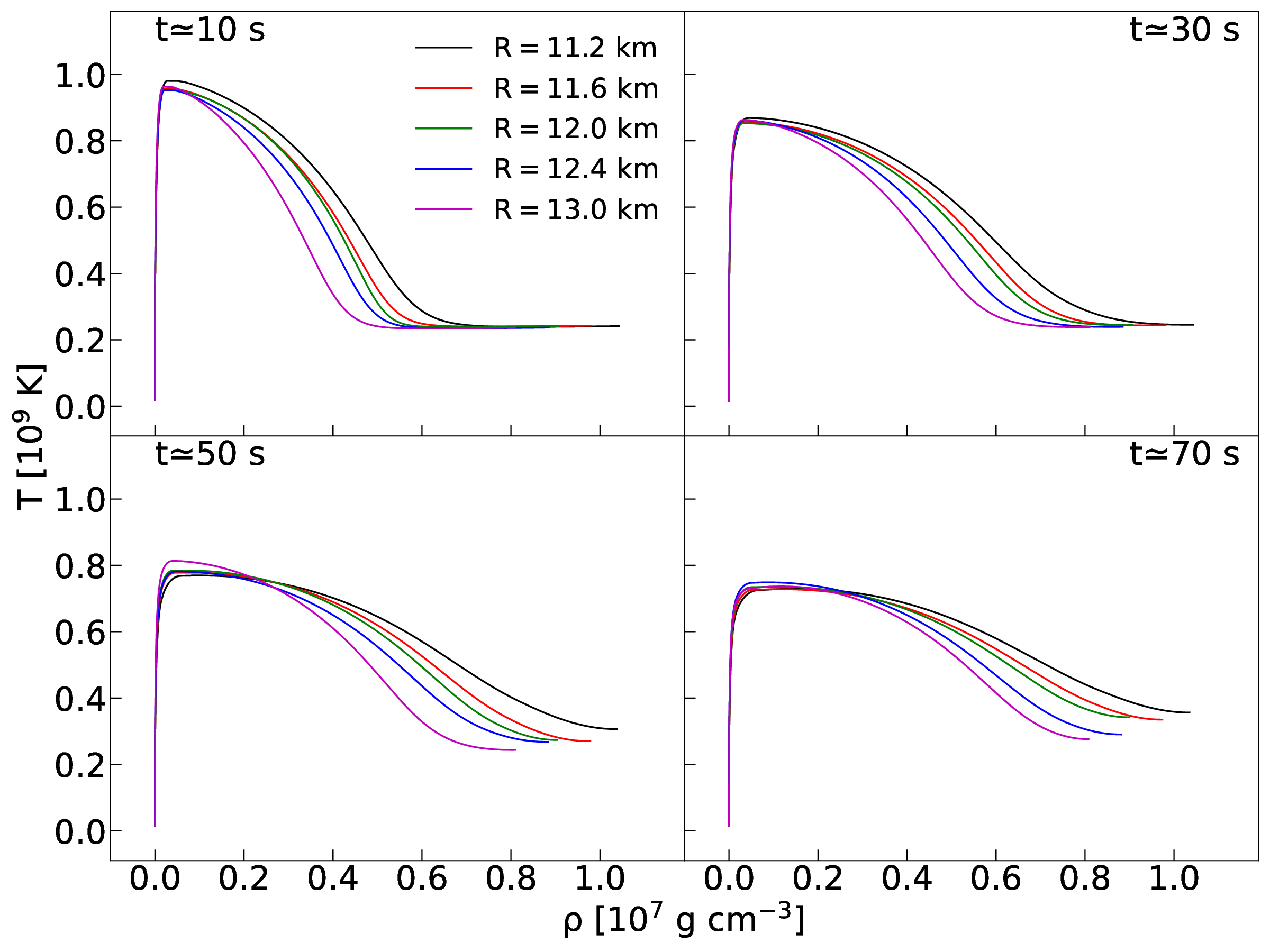} 
\caption{Same as Figure~\ref{fig:9}, but for the cases with different NS radii.
}
\label{fig:10}
\end{figure}

In order to see the dependence of the thermal structure on the type I X-ray burst nucleosynthesis, the temperature-density diagrams are calculated with the variations of NS masses and radii. Results are summarized in Figures~\ref{fig:9} and~\ref{fig:10}, which shows the changes of the structure curves after the peak of the burst during a burst cycle. We can see that high-mass or small-radius models have wider density regions where the temperature is relatively higher, while the low-mass or large-radius models have narrower ones, which leads to the short cooling timescale with smaller mass or larger radius models. Thus, we may infer that the compact NS models can proceed with $rp-$process nucleosynthesis as the NS can be warm for a longer time.

\section{Conclusions} \label{sec:con}
In this work, we perform a composition analysis of the mixed H/He burning ashes in the various input conditions with the code {\ttfamily{MESA}}. The relations between the average mass number of burst ashes ($A_{\rm {ash}}$) and burst strength ($\alpha$), the values of charge number $\langle Z\rangle$ and impurity parameter $Q_{\rm imp}$ are systematically studied. Our principal conclusions from the present work are as follows:

1. The NS structure has pivotal influences on the composition of the burst ashes. The greater the surface gravity, the larger the $A_{\rm ash}$ and $\langle Z\rangle$, but for the higher NS mass case due to the competition between the envelope temperature and the recurrence time. Most of the changes in the composition of the burst ashes are in the $A\sim65-107$ region when altering NS mass, in the $A\sim85-107$ region when altering NS radius.

2. Both the base heating and metallicity have significant effects on the composition of the burst ashes. When changing base heating, the mass fraction of burst ashes mainly changes in the $A\sim85-107$. When changing metallicity, the mass fraction of burst ashes mainly changes in the $A\sim65-107$ region.

3. The new reaction rates from the $(p,\gamma)$ reactions on $^{17}\rm{F}$, $^{19}\rm{F}$, $^{26}\rm{P}$, $^{56}\rm{Cu}$, $^{65}\rm{As}$, and $(\alpha,p)$ reaction on $^{22} \rm{Mg}$ have no discernible impact on the yield of final products.

4. The synthesis of heavy nuclei is positively related to the X-ray bursts having long tails and high peak luminosity.
The largest differences of the mass fraction of burst ashes from our calculations with various models can vary by a factor of $\sim10^6$.

5. The changes of $A_{\rm ash}$ and $\alpha$ are analysed with the variations of NS mass, radius and burst input parameters such as base heating, metallicity and new reaction rates. Both $A_{\rm ash}$ and $\alpha$ increase as mass increases, decrease as radius decreases. Thus, we expect $A_{\rm ash}$ is monotonically related to the NS surface gravity. The ashes from the mixed H/He X-ray burst nucleosynthesis become heavier in more compact NSs.
The effects of metallicity and base heating on $A_{\rm ash}$ and $\alpha$ are similar to the effect of NS radius. The ashes become heavier in lower metallicity and lower base heating scenarios. 

6. The values of $\langle Z\rangle$ and $Q_{\rm{imp}}$ are calculated with the variations of NS mass, radius, base heating and metallicity. $\langle Z\rangle$ increases as mass increases, decreases as radius increases. The changes of $Q_{\rm{imp}}$ are not obvious with the variation of NS mass and radius.  
As base heating increases, $\langle Z\rangle$ decreases, $Q_{\rm{imp}}$ increases. 
As metallicity increases, both the $\langle Z\rangle$ and $Q_{\rm{imp}}$ decreases.  $Q_{\rm{imp}}$ become larger in smaller metallicity and larger base heating.

In our MESA calculation, we treat the NS mass and $Q_b$ as independent parameters, but in reality, they should be correlated due to the presence of neutrino cooling inside the NS core, such as the modified Urca process and baryon bremsstrahlung~\citep{2021ApJ...923...64D}. 
The nuclear reaction network (\textit{rp.net}) adopted in this work does not include the p-nuclei mentioned in the text.
The investigation of NS-mass impacts on burst ashes, considering neutrino cooling inside the core (that is, without any artificial parameters such as $Q_b$) with a p-nuclei included nuclear reaction network, is left for our forthcoming paper.

To sum up, our results provide a systematic analysis of the mixed H/He X-ray burst final products, the NS structure and input parameters (base heating and metallicity) have non-negligible impacts on the composition of the burst ashes, which have significant influences on the NS thermal structure and the observed galactic abundances of light $p$-nuclei.

\section*{Acknowledgement}
The authors are grateful to the anonymous referee for reading and for the instructive comments on our manuscript prior to publication. We thank M. Hashimoto for his encouragement. We thank J. Hu for sharing their new $^{22}{\rm Mg}(\alpha,p)^{25}{\rm Al}$ reaction rate. This work received the generous support of the Natural Science Foundation of Xinjiang no. 2024D01C52, National Natural Science Foundation of China Nos. 12263006, U2031204, 12163005, 12373038, and the Major Science and Technology Program of Xinjiang Uygur Autonomous Region under Grant No. 2022A03013-3. This work was also supported by JSPS KAKENHI (JP20H05648, JP21H01087, JP23K19056, JP24H00008). N.N. was also supported by the RIKEN Intensive Research Project (FY2024--2025).

\textbf{Software and data citations:} \texttt{MESA}(v9793;~\citep{Paxton2011ApJS..192....3P, 2013ApJS..208....4P, Paxton2015ApJS..220...15P, Paxton2018ApJS..234...34P}). Relevant \texttt{MESA} inputs for this work are available on Zenodo\dataset[doi:10.5281/zenodo.14604038]{https://doi.org/10.5281/zenodo.14604038}

\vspace{5mm}





\appendix
\section{X-ray burst models} \label{sec:app}
We show the physical quantities of the burst models in the present work, where the input parameters (i.e. NS mass $M$, radius $R$, metallicity $Z$, base heating $Q_{\rm b}$) and the output parameters (burst strength $\alpha$, burst energy $E_{\rm burst}$, peak luminosity $L_{\rm peak}$, burst interval $\Delta t$, ignition pressure $P_{\rm ign}$, burst rise time $t_{\rm rise}$, burst duration $\tau$, e-folding time $\tau_{\rm e}$, average mass number $A_{\rm ash}$, average charge number $\langle Z \rangle$, impurity parameter $Q_{\rm imp}$, the most abundant elements ($X>10^{-2}$)), the total number of bursts in each model($N_{\rm T}$) and the number of bursts averaged over in each model($N_{\rm A}$) are given.
\begin{table}[h] 
\centering
\tablenum{1}
\caption{Physical quantities of burst models for the \texttt{MESA} code, where the accretion rate is fixed as $\dot{M}=2\times10^{-9}\,\rm{M_{\odot}/yr}$, the initial hydorgen mass fraction is set as $X=0.7$. Errors in the output parameters indicate the $1\sigma$ standard deviation.}
\setlength{\tabcolsep}{3pt}
\begin{tabular}[c]{ccccccccccc}
    \toprule
    Model & M & R & $\rm Z$ & $\rm Q_{b}$ & $\rm \alpha$ & $\rm E_{burst}$ & $\rm L_{peak}$ & $\rm \Delta t$ & $\rm P_{ign}$ & $\rm t_{rise}$ \\ 
    $\rm Number$ & $ M_{\odot}$ & $\rm km$ & $ $ &  &  & $10^{39}~\rm erg$ & $ 10^{38}~\rm erg/s$ & $\rm h$ & $\rm 10^{22}~dyn~cm^{-2}$ & $\rm s$ \\
    \hline
    1 & 1.1 & 11.2 & 0.02 & 0.1 & 47.17$\pm1.61$ & 4.74$\pm0.15$ & 1.24$\pm0.08$ & 3.47$\pm0.07$ & 2.02$\pm0.08$ & 6.63$\pm0.52$ \\
    2 & 1.3 & 11.2 & 0.02 & 0.1 & 57.49$\pm3.42$ & 4.96$\pm0.25$ & 1.71$\pm0.19$ & 3.68$\pm0.12$ & 2.51$\pm0.10$ & 5.85$\pm0.66$ \\
    3 & 1.4 & 11.2 & 0.02 & 0.1 & 63.56$\pm4.72$ & 4.82$\pm0.28$ & 2.01$\pm0.23$ & 3.63$\pm0.14$ & 2.75$\pm0.12$ & 5.73$\pm0.70$ \\     
    4 & 1.5 & 11.2 & 0.02 & 0.1 & 67.96$\pm3.38$ & 4.75$\pm0.19$ & 2.10$\pm0.23$ & 3.55$\pm0.08$ & 2.94$\pm0.10$ & 5.40$\pm0.99$ \\ 
    5 & 1.7 & 11.2 & 0.02 & 0.1 & 78.69$\pm3.19$ & 4.96$\pm0.25$ & 2.63$\pm0.29$ & 3.71$\pm0.10$ & 3.50$\pm0.29$ & 5.14$\pm0.47$ \\ 
    6 & 1.8 & 11.2 & 0.02 & 0.1 & 83.24$\pm5.86$ & 5.19$\pm0.40$ & 2.98$\pm0.28$ & 3.83$\pm0.11$ & 3.90$\pm0.20$ & 4.78$\pm0.79$ \\
    7 & 1.4 & 11.2 & 0.01 & 0.1 & 64.13$\pm3.61$ & 5.47$\pm0.25$ & 1.93$\pm0.34$ & 4.16$\pm0.12$ & 3.08$\pm0.09$ & 5.25$\pm0.67$ \\    
    8 & 1.4 & 11.6 & 0.01 & 0.1 & 61.32$\pm2.50$ & 5.89$\pm0.19$ & 1.82$\pm0.15$ & 4.47$\pm0.09$ & 2.84$\pm0.08$ & 5.58$\pm0.91$ \\ 
    9 & 1.4 & 12.0 & 0.01 & 0.1 & 58.04$\pm3.02$ & 6.47$\pm0.31$ & 1.75$\pm0.12$ & 4.82$\pm0.10$ & 2.63$\pm0.09$ & 6.08$\pm0.68$ \\ 
    10 & 1.4 & 12.4 & 0.01 & 0.1 & 56.15$\pm3.40$ & 6.82$\pm0.42$ & 1.70$\pm0.17$ & 5.09$\pm0.15$ & 2.42$\pm0.08$ & 6.47$\pm0.56$ \\ 
    11 & 1.4 & 13.0 & 0.01 & 0.1 & 52.63$\pm2.76$ & 7.50$\pm0.31$ & 1.63$\pm0.13$ & 5.54$\pm0.15$ & 2.20$\pm0.09$ & 6.69$\pm0.71$ \\
    12 & 1.4 & 11.2 & 0.02 & 0.3 & 58.30$\pm1.78$ & 4.60$\pm0.16$ & 1.46$\pm0.09$ & 3.19$\pm0.06$ & 2.51$\pm0.09$ & 6.05$\pm0.61$ \\
    13 & 1.4 & 11.2 & 0.02 & 0.5 & 54.20$\pm3.02$ & 4.36$\pm0.26$ & 1.22$\pm0.08$ & 2.80$\pm0.05$ & 2.31$\pm0.09$ & 6.24$\pm0.49$ \\
    14 & 1.4 & 11.2 & 0.02 & 0.7 & 51.43$\pm2.27$ & 4.13$\pm0.18$ & 1.07$\pm0.06$ & 2.52$\pm0.05$ & 2.18$\pm0.11$ & 6.24$\pm0.47$ \\
    15 & 1.4 & 11.2 & 0.005 & 0.1 & 64.29$\pm3.51$ & 6.23$\pm0.31$ & 1.88$\pm0.12$ & 4.71$\pm0.12$ & 3.29$\pm0.44$ & 5.52$\pm0.63$ \\
    16 & 1.4 & 11.2 & 0.05 & 0.1 & 59.83$\pm2.68$ & 4.02$\pm0.16$ & 2.11$\pm0.16$ & 2.86$\pm0.06$ & 2.32$\pm0.20$ & 5.64$\pm0.60$ \\ 
    17(R1) & 1.4 & 11.2 & 0.02 & 0.1 & 62.81$\pm3.33$ & 4.75$\pm0.24$ & 1.94$\pm0.25$ & 3.54$\pm0.07$ & 2.70$\pm0.08$ & 5.65$\pm0.97$ \\ 
    18(R2) & 1.4 & 11.2 & 0.02 & 0.1 & 63.12$\pm3.36$ & 4.78$\pm0.18$ & 1.98$\pm0.27$ & 3.58$\pm0.08$ & 2.71$\pm0.11$ & 5.85$\pm0.74$ \\ 
    19(R3) & 1.4 & 11.2 & 0.02 & 0.1 & 62.81$\pm3.42$ & 4.81$\pm0.19$ & 1.88$\pm0.15$ & 3.59$\pm0.08$ & 2.72$\pm0.10$ & 5.86$\pm0.88$ \\
    20(R4) & 1.4 & 11.2 & 0.02 & 0.1 & 61.13$\pm4.99$ & 4.94$\pm0.51$ & 1.99$\pm0.39$ & 3.57$\pm0.07$ & 2.70$\pm0.07$ & 5.94$\pm0.74$ \\
    21(R5) & 1.4 & 11.2 & 0.02 & 0.1 & 61.91$\pm4.42$ & 4.88$\pm0.30$ & 1.94$\pm0.20$ & 3.58$\pm0.08$ & 2.68$\pm0.07$ & 6.16$\pm0.79$ \\
    22(R6) & 1.4 & 11.2 & 0.02 & 0.1 & 62.33$\pm3.01$ & 4.84$\pm0.17$ & 1.97$\pm0.27$ & 3.58$\pm0.08$ & 2.72$\pm0.12$ & 5.95$\pm0.88$ \\
    23(ALL) & 1.4 & 11.2 & 0.02 & 0.1 & 62.54$\pm3.67$ & 4.79$\pm0.25$ & 1.98$\pm0.20$ & 3.56$\pm0.08$ & 2.72$\pm0.12$ & 5.78$\pm1.04$ \\
    
   \toprule
    Model & $\rm \tau$ & $\rm \tau_{e}$ & $A_{\rm ash}$ & $\langle Z\rangle$ & $Q_{\rm{imp}}$ & \multicolumn{4}{c}{$\rm A$}& $N_{\rm A}/N_{\rm T}$\\
    $\rm Number$ & $\rm s$ & $\rm s$ & & & &\multicolumn{4}{c}{$\rm X(A)>10^{-2}$} & \\
    \hline
    1 & 24.08$\pm3.87$ & 43.02$\pm4.27$ & 39.71 & 18.99 & 54.23 &\multicolumn{4}{c}{64,60,32,28,68,65,61,72,63,30,34,56,36,12}&$28/34$ \\
    2 & 18.40$\pm2.59$ & 32.11$\pm3.61$ & 41.32 & 19.75 & 53.46 &\multicolumn{4}{c}{64,60,32,28,68,65,72,61,63,34,36,30}& $27/33$ \\
    3 & 14.18$\pm2.83$ & 25.44$\pm3.62$ & 41.00 & 19.60 & 53.31 &\multicolumn{4}{c}{64,32,60,28,68,65,72,34,61,36,63}&  $25/31$\\    
    4 & 13.57$\pm2.25$ & 23.51$\pm3.15$ & 41.02 & 19.61 & 53.83 &\multicolumn{4}{c}{64,32,60,28,68,65,72,61,34,63,36,30}&  $25/31$\\ 
    5 & 10.48$\pm2.10$ & 17.97$\pm2.65$ & 41.17 & 19.68 & 52.99 &\multicolumn{4}{c}{64,32,60,28,65,72,34,61,36,63,30,56}& $22/28$ \\ 
    6 & 10.40$\pm1.95$ & 16.81$\pm2.51$ & 41.19 & 19.70 & 52.77 &\multicolumn{4}{c}{64,32,60,28,68,65,72,30,34,36,61,63,56,54}& $19/25$ \\ 
    7 & 16.13$\pm3.77$ & 29.61$\pm5.47$ & 43.96 & 21.02 & 62.16 &\multicolumn{4}{c}{64,32,68,60,28,72,65,69,63,66,67,36,34,73,76}& $24/30$ \\   
    8 & 19.95$\pm3.07$ & 35.73$\pm3.92$ & 43.56 & 20.83 & 61.80 &\multicolumn{4}{c}{64,32,68,60,28,72,65,69,63,66,34,67,36,73,61}& $24/30$ \\ 
    9 & 22.74$\pm3.49$ & 40.69$\pm4.15$ & 43.18 & 20.65 & 61.11 &\multicolumn{4}{c}{64,32,60,68,28,72,65,69,63,66,34,61,67,73,36}& $23/29$ \\ 
    10 & 25.82$\pm3.81$ & 46.08$\pm4.81$ & 43.25 & 20.68 & 60.57 &\multicolumn{4}{c}{64,32,60,68,28,72,65,69,63,61,34,66,67,73,36}& $25/31$ \\ 
    11 & 31.26$\pm4.58$ & 55.00$\pm5.83$ & 42.04 & 20.11 & 59.70 &\multicolumn{4}{c}{64,32,60,28,68,72,65,69,63,61,34,66,67,73}& $25/31$ \\
    12 & 18.16$\pm2.40$ & 32.68$\pm2.66$ & 39.69 & 18.98 & 54.38 &\multicolumn{4}{c}{64,60,32,28,68,65,72,63,61,30,34,36}& $28/34$ \\
    13 & 19.50$\pm2.93$ & 34.08$\pm3.45$ & 38.30 & 18.32 & 54.85 &\multicolumn{4}{c}{64,28,60,32,68,65,30,63,61,72,12,34,36,69,66}& $30/36$ \\
    14 & 18.12$\pm3.22$ & 32.58$\pm4.58$ & 37.61 & 18.00 & 54.42 &\multicolumn{4}{c}{64,28,60,32,68,30,65,63,61,12,72,66,67,69,36,34,62,40,54,56}& $32/38$ \\
    15 & 16.48$\pm2.51$ & 30.95$\pm2.99$ & 43.54 & 20.93 & 74.17 &\multicolumn{4}{c}{64,32,68,28,60,72,65,69,73,36,66,76,67,34,82,75,63}& $11/20$ \\
    16 & 11.05$\pm1.18$ & 15.18$\pm1.17$ & 36.73 & 17.62 & 43.67 &\multicolumn{4}{c}{60,28,64,32,56,68,30,34,61,12,36,40,54,52,31}& $31/37$ \\ 
    17(R1) & 16.14$\pm3.05$ & 27.49$\pm3.26$ & 40.02 & 19.15 & 52.66 &\multicolumn{4}{c}{64,60,32,28,68,65,72,34,61,36,56,63,30}& $25/31$ \\
    18(R2) & 14.17$\pm3.03$ & 25.42$\pm3.92$ & 40.81 & 19.51 & 54.15 &\multicolumn{4}{c}{64,32,60,28,68,65,72,34,61,36,63,30}& $26/32$ \\ 
    19(R3) & 15.03$\pm2.46$ & 26.97$\pm2.76$ & 41.37 & 19.78 & 53.54 &\multicolumn{4}{c}{64,60,32,28,68,65,72,34,61,63,36,30}& $26/32$ \\
    20(R4) & 13.45$\pm3.70$ & 24.42$\pm5.46$ & 40.73 & 19.48 & 53.51 &\multicolumn{4}{c}{64,32,60,28,68,65,72,34,61,36,63,56,30}& $25/31$ \\
    21(R5) & 14.17$\pm2.67$ & 25.53$\pm3.55$ & 40.91 & 19.57 & 53.56 &\multicolumn{4}{c}{64,32,60,28,68,65,72,34,61,36,63,30}& $26/32$ \\
    22(R6) & 14.40$\pm2.74$ & 25.59$\pm3.76$ & 40.81 & 19.51 & 53.61 &\multicolumn{4}{c}{64,32,60,28,68,65,72,34,61,56,36,63,30}& $26/32$ \\
    23(ALL) & 16.31$\pm3.79$ & 27.01$\pm3.88$ & 38.80 & 18.61 & 52.66 &\multicolumn{4}{c}{64,60,32,28,68,65,56,72,36,30,61,34,63,54,40}& $19/25$ \\
    \hline
\toprule
\end{tabular}
\tablecomments{ R1:${}^{\rm 22}{\rm Mg}(\alpha,p){}^{\rm 25}{\rm Al}$~\citep{2021PhRvL.127q2701H},R2:${}^{\rm 26}{\rm P}(p,\gamma){}^{\rm 27}{\rm S}$~\citep{2023ApJ...950..133H},R3:${}^{\rm 65}{\rm As}(p,\gamma){}^{\rm 66}{\rm Se}$~\citep{2022ApJ...929...72L},R4:${}^{\rm 19}{\rm F}(p,\gamma){}^{\rm 20}{\rm Ne}$~\citep{2022Natur.610..656Z},R5:${}^{\rm 17}{\rm F}(p,\gamma){}^{\rm 18}{\rm Ne}$~\citep{2017MNRAS.464L...6K},R6:${}^{\rm 57}{\rm Cu}(p,\gamma){}^{\rm 58}{\rm Zn}$~\citep{2022ApJ...929...73L}}
\label{tab:inp}
\end{table}

\begin{table}[h] 
\centering
\tablenum{2}
\caption{The mass number ($A$) and mass fraction ($X$) of the abundant nuclei for the burst ashes in each scenario (mass fraction $>10^{-2}$).}

\renewcommand{\arraystretch}{0.9}
\begin{tabular}[c]{ccccccccccccccccc}
    \toprule
    \multirow{2}{*}{$\rm Ranking$} & \multicolumn{2}{c}{$\rm Model\ 1$} &\multicolumn{2}{c}{$\rm Model\ 2$} & \multicolumn{2}{c}{$\rm Model\ 3$} & \multicolumn{2}{c}{$\rm Model\ 4$} & \multicolumn{2}{c}{$\rm Model\ 5$}&\multicolumn{2}{c}{$\rm Model\ 6$} & \multicolumn{2}{c}{$\rm Model\ 7$} &\multicolumn{2}{c}{$\rm Model\ 8$} \\
     & $\rm A$ & $\rm X$ & $\rm A$ & $\rm X$ & $\rm A$ & $\rm X$ & $\rm A$ & $\rm X$ & $\rm A$ & $\rm X$ & $\rm A$ & $\rm X$ & $\rm A$ & $\rm X$ & $\rm A$ & $\rm X$ \\
    \hline
    1 & 64 & 2.30E-1 & 64  & 2.76E-1 & 64 & 2.91E-1 & 64 & 2.96E-1 & 64 & 3.03E-1 & 64 & 3.07E-1 & 64 & 2.81E-1 & 64  & 2.75E-1 \\
    2 & 60 & 1.64E-1 & 60  & 1.46E-1 & 32 & 1.48E-1 & 32 & 1.41E-1 & 32 & 1.41E-1 & 32 & 1.32E-1 & 32 & 1.37E-1 & 32  & 1.42E-1 \\
    3 & 32 & 1.26E-1 & 32  & 1.41E-1 & 60 & 1.42E-1 & 60 & 1.36E-1 & 60 & 1.32E-1 & 60 & 1.29E-1 & 68 & 9.62E-2 & 68  & 9.37E-2 \\    
    4 & 28 & 1.22E-1 & 28  & 9.53E-2 & 28 & 8.57E-2 & 28 & 9.63E-2 & 28 & 9.30E-2 & 28 & 1.04E-1 & 60 & 7.70E-2 & 60  & 8.41E-2 \\ 
    5 & 68 & 6.17E-2 & 68  & 6.93E-2 & 68 & 6.97E-2 & 68 & 6.77E-2 & 68 & 7.07E-2 & 68 & 7.02E-2 & 28 & 6.68E-2 & 28  & 6.36E-2 \\ 
    6 & 65 & 1.79E-2 & 65  & 1.83E-2 & 65 & 1.72E-2 & 65 & 1.80E-2 & 65 & 1.69E-2 & 65 & 1.74E-2 & 72 & 3.05E-2 & 72  & 2.96E-2 \\ 
    7 & 61 & 1.61E-2 & 72  & 1.69E-2 & 72 & 1.66E-2 & 72 & 1.61E-2 & 72 & 1.64E-2 & 72 & 1.59E-2 & 65 & 2.44E-2 & 65  & 2.45E-2 \\   
    8 & 72 & 1.58E-2 & 61  & 1.42E-2 & 34 & 1.41E-2 & 61 & 1.31E-2 & 34 & 1.48E-2 & 30 & 1.37E-2 & 69 & 1.87E-2 & 69  & 1.85E-2 \\ 
    9 & 63 & 1.53E-2 & 63  & 1.32E-2 & 61 & 1.29E-2 & 34 & 1.30E-2 & 61 & 1.28E-2 & 34 & 1.35E-2 & 63 & 1.31E-2 & 63  & 1.38E-2 \\ 
    10 & 30 & 1.41E-2 & 34 & 1.32E-2 & 36 & 1.24E-2 & 63 & 1.26E-2 & 36 & 1.27E-2 & 36 & 1.28E-2 & 66 & 1.27E-2 & 66 & 1.26E-2 \\ 
    11 & 34 & 1.27E-2 & 36 & 1.18E-2 & 63 & 1.18E-2 & 36 & 1.22E-2 & 63 & 1.16E-2 & 61 & 1.26E-2 & 67 & 1.25E-2 & 34 & 1.24E-2 \\
    12 & 56 & 1.08E-2 & 30 & 1.06E-2 &  &  & 30 & 1.13E-2 & 30 & 1.16E-2 & 63 & 1.07E-2 & 36 & 1.24E-2 & 67 & 1.23E-2 \\
    13 & 36 & 1.08E-2 &  &  &  &  &  &  & 56 & 1.06E-2 & 56 & 1.04E-2 & 34 & 1.19E-2 & 36 & 1.19E-2 \\
    14 & 12 & 1.01E-2 &  &  &  &  &  &  &  &  & 54 & 1.02E-2 & 73 & 1.17E-2 & 73 & 1.17E-2 \\
    15 & & & & & & & & & & & & & 76 & 1.00E-2 & 61 & 1.07E-2 \\
    \toprule
    \multirow{2}{*}{$\rm Ranking$}& \multicolumn{2}{c}{$\rm Model\ 9$} & \multicolumn{2}{c}{$\rm Model\ 10$} & \multicolumn{2}{c}{$\rm Model\ 11$}&\multicolumn{2}{c}{$\rm Model\ 12$} & \multicolumn{2}{c}{$\rm Model\ 13$} &\multicolumn{2}{c}{$\rm Model\ 14$} & \multicolumn{2}{c}{$\rm Model\ 15$} & \multicolumn{2}{c}{$\rm Model\ 16$} \\
     & $\rm A$ & $\rm X$ & $\rm A$ & $\rm X$ & $\rm A$ & $\rm X$ & $\rm A$ & $\rm X$ & $\rm A$ & $\rm X$ & $\rm A$ & $\rm X$ & $\rm A$ & $\rm X$ & $\rm A$ & $\rm X$ \\
    \hline
    1& 64 & 2.68E-1 & 64 & 2.59E-1 & 64 & 2.47E-1 & 64 & 2.59E-1 & 64 & 2.36E-1 & 64  & 2.12E-1 & 64 & 2.68E-1 & 60 & 2.69E-1 \\
    2& 32 & 1.38E-1 & 32 & 1.36E-1 & 32 & 1.32E-1 & 60 & 1.39E-1 & 28 & 1.46E-1 & 28  & 1.66E-1 & 32 & 1.38E-1 & 28 & 1.57E-1 \\
    3& 60 & 9.58E-2 & 60 & 1.03E-1 & 60 & 1.19E-1 & 32 & 1.22E-1 & 60 & 1.41E-1 & 60  & 1.42E-1 & 68 & 9.58E-2 & 64 & 1.52E-1 \\
    4& 68 & 8.74E-2 & 68 & 8.59E-2 & 28 & 9.01E-2 & 28 & 1.20E-1 & 32 & 1.02E-1 & 32  & 8.84E-2 & 28 & 8.74E-2 & 32 & 1.38E-1 \\
    5& 28 & 7.22E-2 & 28 & 7.56E-2 & 68 & 7.76E-2 & 68 & 6.71E-2 & 68 & 6.15E-2 & 68  & 5.79E-2 & 60 & 7.22E-2 & 56 & 2.30E-2 \\
    6& 72 & 2.76E-2 & 72 & 2.69E-2 & 72 & 2.41E-2 & 65 & 1.95E-2 & 65 & 2.04E-2 & 30  & 2.16E-2 & 72 & 2.76E-2 & 68 & 1.91E-2 \\
    7& 65 & 2.33E-2 & 65 & 2.29E-2 & 65 & 2.19E-2 & 72 & 1.61E-2 & 30 & 1.90E-2 & 65  & 2.09E-2 & 65 & 2.33E-2 & 30 & 1.87E-2 \\
    8& 69 & 1.79E-2 & 69 & 1.81E-2 & 69 & 1.68E-2 & 63 & 1.58E-2 & 63 & 1.85E-2 & 63  & 2.05E-2 & 69 & 1.79E-2 & 34 & 1.69E-2 \\
    9& 63 & 1.47E-2 & 63 & 1.55E-2 & 63 & 1.67E-2 & 61 & 1.55E-2 & 61 & 1.69E-2 & 61  & 1.84E-2 & 73 & 1.47E-2 & 61 & 1.33E-2 \\
    10& 66 & 1.21E-2 & 61 & 1.28E-2 & 61 & 1.44E-2 & 30 & 1.46E-2 & 72 & 1.46E-2 & 12 & 1.63E-2 & 36 & 1.21E-2 & 12 & 1.31E-2 \\
    11& 34 & 1.18E-2 & 34 & 1.26E-2 & 34 & 1.21E-2 & 34 & 1.32E-2 & 12 & 1.26E-2 & 72 & 1.39E-2 & 66 & 1.18E-2 & 36 & 1.19E-2 \\
    12& 61 & 1.18E-2 & 66 & 1.23E-2 & 66 & 1.17E-2 & 36 & 1.10E-2 & 34 & 1.17E-2 & 66 & 1.11E-2 & 76 & 1.18E-2 & 40 & 1.14E-2 \\
    13& 67 & 1.17E-2 & 67 & 1.22E-2 & 67 & 1.13E-2 &  &  & 36 & 1.06E-2 & 67 & 1.10E-2 & 67 & 1.17E-2 & 54 & 1.05E-2 \\
    14& 73 & 1.13E-2 & 73 & 1.14E-2 & 73 & 1.04E-2 &  &  & 69 & 1.03E-2 & 69 & 1.07E-2 & 34 & 1.13E-2 & 52 & 1.00E-2 \\
    15& 36 & 1.11E-2 & 36 & 1.04E-2 &  &  &  &  & 66 & 1.00E-2 & 36 & 1.07E-2 & 82 & 1.11E-2 & 31 & 1.00E-2 \\
    16&  &  &  &  &  &  &  &  &  &  & 34 & 1.07E-2 & 75 & 1.11E-2 &  &  \\
    17&  &  &  &  &  &  &  &  &  &  & 62 & 1.06E-2 & 63 & 1.11E-2 &  &  \\
    18&  &  &  &  &  &  &  &  &  &  & 40 & 1.04E-2 &  &  &  &  \\
    19&  &  &  &  &  &  &  &  &  &  & 54 & 1.01E-2 &  &  &  &  \\
    20&  &  &  &  &  &  &  &  &  &  & 56 & 1.00E-2 &  &  &  &  \\
    \toprule
    \multirow{2}{*}{$\rm Ranking$}& \multicolumn{2}{c}{$\rm Model\ 17$}&\multicolumn{2}{c}{$\rm Model\ 18$}  & \multicolumn{2}{c}{$\rm Model\ 19$} &\multicolumn{2}{c}{$\rm Model\ 20$} & \multicolumn{2}{c}{$\rm Model\ 21$} & \multicolumn{2}{c}{$\rm Model\ 22$} & \multicolumn{2}{c}{$\rm Model\ 23$}&\multicolumn{2}{c}{} \\
    & $\rm A$ & $\rm X$ & $\rm A$ & $\rm X$ & $\rm A$ & $\rm X$ & $\rm A$ & $\rm X$ & $\rm A$ & $\rm X$ & $\rm A$ & $\rm X$ & $\rm A$ & $\rm X$ &  &  \\
    \hline
    1 & 64 & 2.83E-1 & 64 & 2.90E-1 & 64 & 2.88E-1 & 64  & 2.85E-1 & 64 & 2.90E-1 & 64 & 2.92E-1 & 64 & 2.74E-1 &  &  \\
    2 & 60 & 1.51E-1 & 32 & 1.41E-1 & 60 & 1.42E-1 & 32  & 1.48E-1 & 32 & 1.45E-1 & 32 & 1.43E-1 & 60 & 1.43E-1 &  & \\
    3 & 32 & 1.45E-1 & 60 & 1.41E-1 & 32 & 1.40E-1 & 60  & 1.42E-1 & 60 & 1.43E-2 & 60 & 1.41E-1 & 32 & 1.43E-1 &  & \\    
    4 & 28 & 9.76E-2 & 28 & 9.21E-2 & 28 & 9.14E-2 & 28  & 8.54E-2 & 28 & 8.81E-2 & 28 & 8.71E-2 & 28 & 1.17E-1 &  & \\ 
    5 & 68 & 6.29E-2 & 68 & 6.92E-2 & 68 & 7.15E-2 & 68  & 6.92E-2 & 68 & 6.90E-2 & 68 & 6.93E-2 & 68 & 5.99E-2 &  &  \\ 
    6 & 65 & 1.64E-2 & 65 & 1.82E-2 & 65 & 1.75E-2 & 65  & 1.79E-2 & 65 & 1.78E-2 & 65 & 1.77E-2 & 65 & 1.61E-2 &  & \\ 
    7 & 72 & 1.47E-2 & 72 & 1.67E-2 & 72 & 1.72E-2 & 72  & 1.68E-2 & 72 & 1.66E-2 & 72 & 1.67E-2 & 56 & 1.45E-2 &  & \\   
    8 & 34 & 1.35E-2 & 34 & 1.31E-2 & 34 & 1.35E-2 & 34  & 1.45E-2 & 34 & 1.35E-2 & 34 & 1.34E-2 & 72 & 1.40E-2 &  & \\ 
    9 & 61 & 1.27E-2 & 61 & 1.30E-2 & 61 & 1.28E-2 & 61  & 1.30E-2 & 61 & 1.29E-2 & 61 & 1.30E-2 & 36 & 1.36E-2 &  & \\ 
    10 & 36 & 1.25E-2 & 36 & 1.23E-2 & 63 & 1.22E-2 & 36 & 1.24E-2 & 36 & 1.22E-2 & 56 & 1.24E-2 & 30 & 1.30E-2 &  & \\ 
    11 & 56 & 1.19E-2 & 63 & 1.21E-2 & 36 & 1.21E-2 & 63 & 1.24E-2 & 63 & 1.21E-2 & 36 & 1.22E-2 & 61 & 1.23E-2 &  & \\
    12 & 63 & 1.17E-2 & 30 & 1.10E-2 & 30 & 1.08E-2 & 56 & 1.05E-2 & 30 & 1.05E-2 & 63 & 1.21E-2 & 34 & 1.22E-2 &  & \\
    13 & 30 & 1.12E-2 &  &  &  &  & 30 & 1.02E-2 &  &  & 30 & 1.01E-2 & 63 & 1.08E-2 &  & \\
    14 &  &  &  &  &  &  &  &  &  &  &  &  & 54 & 1.01E-2 &  &  \\
    15 &  &  &  &  &  &  &  &  &  &  &  &  & 40 & 1.01E-2 &  &  \\    
\toprule
\end{tabular}
\label{tab:AX}
\end{table}

\bibliography{composition}{}

\begin{thebibliography}{}
\expandafter\ifx\csname natexlab\endcsname\relax\def\natexlab#1{#1}\fi
\providecommand{\url}[1]{\href{#1}{#1}}
\providecommand{\dodoi}[1]{doi:~\href{http://doi.org/#1}{\nolinkurl{#1}}}
\providecommand{\doeprint}[1]{\href{http://ascl.net/#1}{\nolinkurl{http://ascl.net/#1}}}
\providecommand{\doarXiv}[1]{\href{https://arxiv.org/abs/#1}{\nolinkurl{https://arxiv.org/abs/#1}}}

\bibitem[{{Alizai} {et~al.}(2023){Alizai}, {Chenevez}, {Cumming}, {Degenaar}, {Falanga}, {Galloway}, {in't Zand}, {Jaisawal}, {Keek}, {Kuulkers}, {Lampe}, {Schatz}, \& {Serino}}]{2023MNRAS.521.3608A}
{Alizai}, K., {Chenevez}, J., {Cumming}, A., {et~al.} 2023, \mnras, 521, 3608, \dodoi{10.1093/mnras/stad374}

\bibitem[{{Bildsten}(1998)}]{1998ASIC..515..419B}
{Bildsten}, L. 1998, in NATO Advanced Study Institute (ASI) Series C, Vol. 515, The Many Faces of Neutron Stars., ed. R.~{Buccheri}, J.~{van Paradijs}, \& A.~{Alpar}, 419, \dodoi{10.48550/arXiv.astro-ph/9709094}

\bibitem[{{Brown} \& {Cumming}(2009)}]{2009ApJ...698.1020B}
{Brown}, E.~F., \& {Cumming}, A. 2009, \apj, 698, 1020, \dodoi{10.1088/0004-637X/698/2/1020}

\bibitem[{{Burgio} {et~al.}(2021){Burgio}, {Schulze}, {Vida{\~n}a}, \& {Wei}}]{2021PrPNP.12003879B}
{Burgio}, G.~F., {Schulze}, H.~J., {Vida{\~n}a}, I., \& {Wei}, J.~B. 2021, Progress in Particle and Nuclear Physics, 120, 103879, \dodoi{10.1016/j.ppnp.2021.103879}

\bibitem[{{Cornelisse} {et~al.}(2000){Cornelisse}, {Heise}, {Kuulkers}, {Verbunt}, \& {in't Zand}}]{2000A&A...357L..21C}
{Cornelisse}, R., {Heise}, J., {Kuulkers}, E., {Verbunt}, F., \& {in't Zand}, J.~J.~M. 2000, \aap, 357, L21, \dodoi{10.48550/arXiv.astro-ph/0003454}

\bibitem[{{Cyburt} {et~al.}(2016){Cyburt}, {Amthor}, {Heger}, {Johnson}, {Keek}, {Meisel}, {Schatz}, \& {Smith}}]{2016ApJ...830...55C}
{Cyburt}, R.~H., {Amthor}, A.~M., {Heger}, A., {et~al.} 2016, \apj, 830, 55, \dodoi{10.3847/0004-637X/830/2/55}

\bibitem[{{Deibel} {et~al.}(2016){Deibel}, {Meisel}, {Schatz}, {Brown}, \& {Cumming}}]{2016ApJ...831...13D}
{Deibel}, A., {Meisel}, Z., {Schatz}, H., {Brown}, E.~F., \& {Cumming}, A. 2016, \apj, 831, 13, \dodoi{10.3847/0004-637X/831/1/13}

\bibitem[{{Dohi} {et~al.}(2020){Dohi}, {Hashimoto}, {Yamada}, {Matsuo}, \& {Fujimoto}}]{2020PTEP.2020c3E02D}
{Dohi}, A., {Hashimoto}, M.-a., {Yamada}, R., {Matsuo}, Y., \& {Fujimoto}, M.~Y. 2020, Progress of Theoretical and Experimental Physics, 2020, 033E02, \dodoi{10.1093/ptep/ptaa010}

\bibitem[{{Dohi} {et~al.}(2024){Dohi}, {Iwakiri}, {Nishimura}, {Noda}, {Nagataki}, \& {Hashimoto}}]{2024ApJ...960...14D}
{Dohi}, A., {Iwakiri}, W.~B., {Nishimura}, N., {et~al.} 2024, \apj, 960, 14, \dodoi{10.3847/1538-4357/ad0a67}

\bibitem[{{Dohi} {et~al.}(2021){Dohi}, {Nishimura}, {Hashimoto}, {Matsuo}, {Noda}, \& {Nagataki}}]{2021ApJ...923...64D}
{Dohi}, A., {Nishimura}, N., {Hashimoto}, M., {et~al.} 2021, \apj, 923, 64, \dodoi{10.3847/1538-4357/ac2821}

\bibitem[{{Dohi} {et~al.}(2022){Dohi}, {Nishimura}, {Sotani}, {Noda}, {Liu}, {Nagataki}, \& {Hashimoto}}]{2022ApJ...937..124D}
{Dohi}, A., {Nishimura}, N., {Sotani}, H., {et~al.} 2022, \apj, 937, 124, \dodoi{10.3847/1538-4357/ac8dfe}

\bibitem[{{Fisker} {et~al.}(2008){Fisker}, {Schatz}, \& {Thielemann}}]{2008ApJS..174..261F}
{Fisker}, J.~L., {Schatz}, H., \& {Thielemann}, F.-K. 2008, \apjs, 174, 261, \dodoi{10.1086/521104}

\bibitem[{{Fisker} {et~al.}(2004){Fisker}, {Thielemann}, \& {Wiescher}}]{2004ApJ...608L..61F}
{Fisker}, J.~L., {Thielemann}, F.-K., \& {Wiescher}, M. 2004, \apjl, 608, L61, \dodoi{10.1086/422215}

\bibitem[{{Fujimoto} {et~al.}(1981){Fujimoto}, {Hanawa}, \& {Miyaji}}]{1981ApJ...247..267F}
{Fujimoto}, M.~Y., {Hanawa}, T., \& {Miyaji}, S. 1981, \apj, 247, 267, \dodoi{10.1086/159034}

\bibitem[{{Galloway} \& {Keek}(2021)}]{2021ASSL..461..209G}
{Galloway}, D.~K., \& {Keek}, L. 2021, in Astrophysics and Space Science Library, Vol. 461, Timing Neutron Stars: Pulsations, Oscillations and Explosions, ed. T.~M. {Belloni}, M.~{M{\'e}ndez}, \& C.~{Zhang}, 209--262, \dodoi{10.1007/978-3-662-62110-3_5}

\bibitem[{{Grevesse} \& {Sauval}(1998)}]{1998SSRv...85..161G}
{Grevesse}, N., \& {Sauval}, A.~J. 1998, \ssr, 85, 161, \dodoi{10.1023/A:1005161325181}

\bibitem[{{Heger} {et~al.}(2007){Heger}, {Cumming}, {Galloway}, \& {Woosley}}]{2007ApJ...671L.141H}
{Heger}, A., {Cumming}, A., {Galloway}, D.~K., \& {Woosley}, S.~E. 2007, \apjl, 671, L141, \dodoi{10.1086/525522}

\bibitem[{{Herrera} {et~al.}(2023){Herrera}, {Sala}, \& {Jos{\'e}}}]{2023A&A...678A.156H}
{Herrera}, Y., {Sala}, G., \& {Jos{\'e}}, J. 2023, \aap, 678, A156, \dodoi{10.1051/0004-6361/202346190}

\bibitem[{{Hou} {et~al.}(2023){Hou}, {Liu}, {Trueman}, {Li}, {Pignatari}, {Bertulani}, \& {Xu}}]{2023ApJ...950..133H}
{Hou}, S.~Q., {Liu}, J.~B., {Trueman}, T.~C.~L., {et~al.} 2023, \apj, 950, 133, \dodoi{10.3847/1538-4357/accf9c}

\bibitem[{{Howard} {et~al.}(1991){Howard}, {Meyer}, \& {Woosley}}]{1991ApJ...373L...5H}
{Howard}, W.~M., {Meyer}, B.~S., \& {Woosley}, S.~E. 1991, \apjl, 373, L5, \dodoi{10.1086/186038}

\bibitem[{{Hu} {et~al.}(2021){Hu}, {Yamaguchi}, {Lam}, {Heger}, {Kahl}, {Jacobs}, {Johnston}, {Xu}, {Zhang}, {Ma}, {Ru}, {Liu}, {Liu}, {Hayakawa}, {Yang}, {Shimizu}, {Hamill}, {Murphy}, {Su}, {Fang}, {Chae}, {Kwag}, {Cha}, {Duy}, {Uyen}, {Kim}, {Pizzone}, {La Cognata}, {Cherubini}, {Romano}, {Tumino}, {Liang}, {Psaltis}, {Sferrazza}, {Kim}, {Li}, \& {Kubono}}]{2021PhRvL.127q2701H}
{Hu}, J., {Yamaguchi}, H., {Lam}, Y.~H., {et~al.} 2021, \prl, 127, 172701, \dodoi{10.1103/PhysRevLett.127.172701}

\bibitem[{{in't Zand} {et~al.}(2011){in't Zand}, {Galloway}, \& {Ballantyne}}]{2011A&A...525A.111I}
{in't Zand}, J.~J.~M., {Galloway}, D.~K., \& {Ballantyne}, D.~R. 2011, \aap, 525, A111, \dodoi{10.1051/0004-6361/201015556}

\bibitem[{{in't Zand} \& {Weinberg}(2010)}]{2010A&A...520A..81I}
{in't Zand}, J.~J.~M., \& {Weinberg}, N.~N. 2010, \aap, 520, A81, \dodoi{10.1051/0004-6361/200913952}

\bibitem[{{Iwakiri} {et~al.}(2021){Iwakiri}, {Serino}, {Mihara}, {Gu}, {Yamaguchi}, {Shidatsu}, \& {Makishima}}]{2021PASJ...73.1405I}
{Iwakiri}, W.~B., {Serino}, M., {Mihara}, T., {et~al.} 2021, \pasj, 73, 1405, \dodoi{10.1093/pasj/psab085}

\bibitem[{{Johnston} {et~al.}(2020){Johnston}, {Heger}, \& {Galloway}}]{2020MNRAS.494.4576J}
{Johnston}, Z., {Heger}, A., \& {Galloway}, D.~K. 2020, \mnras, 494, 4576, \dodoi{10.1093/mnras/staa1054}

\bibitem[{{Jos{\'e}} {et~al.}(2010){Jos{\'e}}, {Moreno}, {Parikh}, \& {Iliadis}}]{2010ApJS..189..204J}
{Jos{\'e}}, J., {Moreno}, F., {Parikh}, A., \& {Iliadis}, C. 2010, \apjs, 189, 204, \dodoi{10.1088/0067-0049/189/1/204}

\bibitem[{{Kajava} {et~al.}(2017){Kajava}, {N{\"a}ttil{\"a}}, {Poutanen}, {Cumming}, {Suleimanov}, \& {Kuulkers}}]{2017MNRAS.464L...6K}
{Kajava}, J.~J.~E., {N{\"a}ttil{\"a}}, J., {Poutanen}, J., {et~al.} 2017, \mnras, 464, L6, \dodoi{10.1093/mnrasl/slw167}

\bibitem[{{Keek} \& {Heger}(2017)}]{2017ApJ...842..113K}
{Keek}, L., \& {Heger}, A. 2017, \apj, 842, 113, \dodoi{10.3847/1538-4357/aa7748}

\bibitem[{{Koike} {et~al.}(1999){Koike}, {Hashimoto}, {Arai}, \& {Wanajo}}]{1999A&A...342..464K}
{Koike}, O., {Hashimoto}, M., {Arai}, K., \& {Wanajo}, S. 1999, \aap, 342, 464

\bibitem[{{Koike} {et~al.}(2004){Koike}, {Hashimoto}, {Kuromizu}, \& {Fujimoto}}]{2004ApJ...603..242K}
{Koike}, O., {Hashimoto}, M.-a., {Kuromizu}, R., \& {Fujimoto}, S.-i. 2004, \apj, 603, 242, \dodoi{10.1086/381354}

\bibitem[{{Kumar} {et~al.}(2024){Kumar}, {Dexheimer}, {Jahan}, {Noronha}, {Noronha-Hostler}, {Ratti}, {Yunes}, {Nava Acuna}, {Alford}, {Anik}, {Chatterjee}, {Chatziioannou}, {Chen}, {Clevinger}, {Conde}, {Cruz-Camacho}, {Dore}, {Drischler}, {Elfner}, {Essick}, {Friedenberg}, {Ghosh}, {Grefa}, {Haas}, {Haber}, {Hammelmann}, {Harris}, {Haster}, {Hatsuda}, {Hippert}, {Hirayama}, {Holt}, {Kahangirwe}, {Karthein}, {Kojo}, {Landry}, {Lin}, {Luzum}, {Manning}, {Salinas San Martin}, {Miller}, {Most}, {Mroczek}, {Muronga}, {Patino}, {Peterson}, {Plumberg}, {Price}, {Providencia}, {Rougemont}, {Roy}, {Shah}, {Shapiro}, {Steiner}, {Strickland}, {Tan}, {Togashi}, {Portillo Vazquez}, {Wen}, {Zhang}, \& {Muses Collaboration}}]{2024LRR....27....3K}
{Kumar}, R., {Dexheimer}, V., {Jahan}, J., {et~al.} 2024, Living Reviews in Relativity, 27, 3, \dodoi{10.1007/s41114-024-00049-6}

\bibitem[{{Kuvin} {et~al.}(2017){Kuvin}, {Belarge}, {Baby}, {Baker}, {Wiedenh{\"o}ver}, {H{\"o}flich}, {Volya}, {Blackmon}, {Deibel}, {Gardiner}, {Lai}, {Linhardt}, {Macon}, {Rasco}, {Quails}, {Colbert}, {Gay}, \& {Keeley}}]{2017PhRvC..96d5812K}
{Kuvin}, S.~A., {Belarge}, J., {Baby}, L.~T., {et~al.} 2017, \prc, 96, 045812, \dodoi{10.1103/PhysRevC.96.045812}

\bibitem[{{Lam} {et~al.}(2022{\natexlab{a}}){Lam}, {Liu}, {Heger}, {Lu}, {Jacobs}, \& {Johnston}}]{2022ApJ...929...72L}
{Lam}, Y.~H., {Liu}, Z.~X., {Heger}, A., {et~al.} 2022{\natexlab{a}}, \apj, 929, 72, \dodoi{10.3847/1538-4357/ac4d8b}

\bibitem[{{Lam} {et~al.}(2022{\natexlab{b}}){Lam}, {Lu}, {Heger}, {Jacobs}, {Smirnova}, {Nieto}, {Johnston}, \& {Kubono}}]{2022ApJ...929...73L}
{Lam}, Y.~H., {Lu}, N., {Heger}, A., {et~al.} 2022{\natexlab{b}}, \apj, 929, 73, \dodoi{10.3847/1538-4357/ac4d89}

\bibitem[{{Lampe} {et~al.}(2016){Lampe}, {Heger}, \& {Galloway}}]{2016ApJ...819...46L}
{Lampe}, N., {Heger}, A., \& {Galloway}, D.~K. 2016, \apj, 819, 46, \dodoi{10.3847/0004-637X/819/1/46}

\bibitem[{{Lau} {et~al.}(2018){Lau}, {Beard}, {Gupta}, {Schatz}, {Afanasjev}, {Brown}, {Deibel}, {Gasques}, {Hitt}, {Hix}, {Keek}, {M{\"o}ller}, {Shternin}, {Steiner}, {Wiescher}, \& {Xu}}]{2018ApJ...859...62L}
{Lau}, R., {Beard}, M., {Gupta}, S.~S., {et~al.} 2018, \apj, 859, 62, \dodoi{10.3847/1538-4357/aabfe0}

\bibitem[{{Lewin} {et~al.}(1993){Lewin}, {van Paradijs}, \& {Taam}}]{1993SSRv...62..223L}
{Lewin}, W. H.~G., {van Paradijs}, J., \& {Taam}, R.~E. 1993, \ssr, 62, 223, \dodoi{10.1007/BF00196124}

\bibitem[{{Meisel}(2018)}]{2018ApJ...860..147M}
{Meisel}, Z. 2018, \apj, 860, 147, \dodoi{10.3847/1538-4357/aac3d3}

\bibitem[{{Meisel} \& {Deibel}(2017)}]{2017ApJ...837...73M}
{Meisel}, Z., \& {Deibel}, A. 2017, \apj, 837, 73, \dodoi{10.3847/1538-4357/aa618d}

\bibitem[{{Meisel} {et~al.}(2018){Meisel}, {Deibel}, {Keek}, {Shternin}, \& {Elfritz}}]{2018JPhG...45i3001M}
{Meisel}, Z., {Deibel}, A., {Keek}, L., {Shternin}, P., \& {Elfritz}, J. 2018, Journal of Physics G Nuclear Physics, 45, 093001, \dodoi{10.1088/1361-6471/aad171}

\bibitem[{{Meisel} {et~al.}(2019){Meisel}, {Merz}, \& {Medvid}}]{2019ApJ...872...84M}
{Meisel}, Z., {Merz}, G., \& {Medvid}, S. 2019, \apj, 872, 84, \dodoi{10.3847/1538-4357/aafede}

\bibitem[{{Meisel} {et~al.}(2015){Meisel}, {George}, {Ahn}, {Bazin}, {Brown}, {Browne}, {Carpino}, {Chung}, {Cole}, {Cyburt}, {Estrad{\'e}}, {Famiano}, {Gade}, {Langer}, {Mato{\v{s}}}, {Mittig}, {Montes}, {Morrissey}, {Pereira}, {Schatz}, {Schatz}, {Scott}, {Shapira}, {Smith}, {Stevens}, {Tan}, {Tarasov}, {Towers}, {Wimmer}, {Winkelbauer}, {Yurkon}, \& {Zegers}}]{2015PhRvL.115p2501M}
{Meisel}, Z., {George}, S., {Ahn}, S., {et~al.} 2015, \prl, 115, 162501, \dodoi{10.1103/PhysRevLett.115.162501}

\bibitem[{{Nava-Callejas} {et~al.}(2024){Nava-Callejas}, {Cavecchi}, \& {Page}}]{2024arXiv241109843N}
{Nava-Callejas}, M., {Cavecchi}, Y., \& {Page}, D. 2024, arXiv e-prints, arXiv:2411.09843, \dodoi{10.48550/arXiv.2411.09843}

\bibitem[{{Parikh} {et~al.}(2008){Parikh}, {Jos{\'e}}, {Moreno}, \& {Iliadis}}]{2008ApJS..178..110P}
{Parikh}, A., {Jos{\'e}}, J., {Moreno}, F., \& {Iliadis}, C. 2008, \apjs, 178, 110, \dodoi{10.1086/589879}

\bibitem[{{Paxton} {et~al.}(2011){Paxton}, {Bildsten}, {Dotter}, {Herwig}, {Lesaffre}, \& {Timmes}}]{Paxton2011ApJS..192....3P}
{Paxton}, B., {Bildsten}, L., {Dotter}, A., {et~al.} 2011, \apjs, 192, 3, \dodoi{10.1088/0067-0049/192/1/3}

\bibitem[{{Paxton} {et~al.}(2013){Paxton}, {Cantiello}, {Arras}, {Bildsten}, {Brown}, {Dotter}, {Mankovich}, {Montgomery}, {Stello}, {Timmes}, \& {Townsend}}]{2013ApJS..208....4P}
{Paxton}, B., {Cantiello}, M., {Arras}, P., {et~al.} 2013, \apjs, 208, 4, \dodoi{10.1088/0067-0049/208/1/4}

\bibitem[{{Paxton} {et~al.}(2015){Paxton}, {Marchant}, {Schwab}, {Bauer}, {Bildsten}, {Cantiello}, {Dessart}, {Farmer}, {Hu}, {Langer}, {Townsend}, {Townsley}, \& {Timmes}}]{Paxton2015ApJS..220...15P}
{Paxton}, B., {Marchant}, P., {Schwab}, J., {et~al.} 2015, \apjs, 220, 15, \dodoi{10.1088/0067-0049/220/1/15}

\bibitem[{{Paxton} {et~al.}(2018){Paxton}, {Schwab}, {Bauer}, {Bildsten}, {Blinnikov}, {Duffell}, {Farmer}, {Goldberg}, {Marchant}, {Sorokina}, {Thoul}, {Townsend}, \& {Timmes}}]{Paxton2018ApJS..234...34P}
{Paxton}, B., {Schwab}, J., {Bauer}, E.~B., {et~al.} 2018, \apjs, 234, 34, \dodoi{10.3847/1538-4365/aaa5a8}

\bibitem[{{Potekhin} \& {Chabrier}(2021)}]{2021A&A...645A.102P}
{Potekhin}, A.~Y., \& {Chabrier}, G. 2021, \aap, 645, A102, \dodoi{10.1051/0004-6361/202039006}

\bibitem[{{Rayet} {et~al.}(1995){Rayet}, {Arnould}, {Hashimoto}, {Prantzos}, \& {Nomoto}}]{1995A&A...298..517R}
{Rayet}, M., {Arnould}, M., {Hashimoto}, M., {Prantzos}, N., \& {Nomoto}, K. 1995, \aap, 298, 517

\bibitem[{{Schatz} {et~al.}(1999){Schatz}, {Bildsten}, {Cumming}, \& {Wiescher}}]{1999ApJ...524.1014S}
{Schatz}, H., {Bildsten}, L., {Cumming}, A., \& {Wiescher}, M. 1999, \apj, 524, 1014, \dodoi{10.1086/307837}

\bibitem[{{Schatz} \& {Rehm}(2006)}]{2006NuPhA.777..601S}
{Schatz}, H., \& {Rehm}, K.~E. 2006, \nphysa, 777, 601, \dodoi{10.1016/j.nuclphysa.2005.05.200}

\bibitem[{{Schatz} {et~al.}(1998){Schatz}, {Aprahamian}, {Goerres}, {Wiescher}, {Rauscher}, {Rembges}, {Thielemann}, {Pfeiffer}, {Moeller}, {Kratz}, {Herndl}, {Brown}, \& {Rebel}}]{1998PhR...294..167S}
{Schatz}, H., {Aprahamian}, A., {Goerres}, J., {et~al.} 1998, \physrep, 294, 167, \dodoi{10.1016/S0370-1573(97)00048-3}

\bibitem[{{Schatz} {et~al.}(2001){Schatz}, {Aprahamian}, {Barnard}, {Bildsten}, {Cumming}, {Ouellette}, {Rauscher}, {Thielemann}, \& {Wiescher}}]{2001PhRvL..86.3471S}
{Schatz}, H., {Aprahamian}, A., {Barnard}, V., {et~al.} 2001, \prl, 86, 3471, \dodoi{10.1103/PhysRevLett.86.3471}

\bibitem[{{Shchechilin} {et~al.}(2021){Shchechilin}, {Gusakov}, \& {Chugunov}}]{2021MNRAS.507.3860S}
{Shchechilin}, N.~N., {Gusakov}, M.~E., \& {Chugunov}, A.~I. 2021, \mnras, 507, 3860, \dodoi{10.1093/mnras/stab2415}

\bibitem[{{Shchechilin} {et~al.}(2023){Shchechilin}, {Gusakov}, \& {Chugunov}}]{2023MNRAS.523.4643S}
---. 2023, \mnras, 523, 4643, \dodoi{10.1093/mnras/stad1731}

\bibitem[{{Song} {et~al.}(2024){Song}, {Liu}, {Zhu}, {Zhen}, {L{\"u}}, \& {Xu}}]{2024MNRAS.529.3103S}
{Song}, L., {Liu}, H., {Zhu}, C., {et~al.} 2024, \mnras, 529, 3103, \dodoi{10.1093/mnras/stae709}

\bibitem[{{Strohmayer} \& {Brown}(2002)}]{2002ApJ...566.1045S}
{Strohmayer}, T.~E., \& {Brown}, E.~F. 2002, \apj, 566, 1045, \dodoi{10.1086/338337}

\bibitem[{{Taam}(1980)}]{1980ApJ...241..358T}
{Taam}, R.~E. 1980, \apj, 241, 358, \dodoi{10.1086/158348}

\bibitem[{{Wallace} \& {Woosley}(1981)}]{1981ApJS...45..389W}
{Wallace}, R.~K., \& {Woosley}, S.~E. 1981, \apjs, 45, 389, \dodoi{10.1086/190717}

\bibitem[{{Wang} {et~al.}(2021){Wang}, {Tan}, {Li}, {Misch}, \& {Sun}}]{2021PhRvL.127q2702W}
{Wang}, L.-J., {Tan}, L., {Li}, Z., {Misch}, G.~W., \& {Sun}, Y. 2021, \prl, 127, 172702, \dodoi{10.1103/PhysRevLett.127.172702}

\bibitem[{{Wijnands} {et~al.}(2017){Wijnands}, {Degenaar}, \& {Page}}]{2017JApA...38...49W}
{Wijnands}, R., {Degenaar}, N., \& {Page}, D. 2017, Journal of Astrophysics and Astronomy, 38, 49, \dodoi{10.1007/s12036-017-9466-5}

\bibitem[{{Woosley} {et~al.}(2004){Woosley}, {Heger}, {Cumming}, {Hoffman}, {Pruet}, {Rauscher}, {Fisker}, {Schatz}, {Brown}, \& {Wiescher}}]{2004ApJS..151...75W}
{Woosley}, S.~E., {Heger}, A., {Cumming}, A., {et~al.} 2004, \apjs, 151, 75, \dodoi{10.1086/381533}

\bibitem[{{Zhang} {et~al.}(2022){Zhang}, {He}, {deBoer}, {Wiescher}, {Heger}, {Kahl}, {Su}, {Odell}, {Chen}, {Li}, {Wang}, {Zhang}, {Cao}, {Zhang}, {Zhang}, {Jiang}, {Wang}, {Li}, {Song}, {Zhao}, {Sun}, {Wu}, {Li}, {Cui}, {Chen}, {Ma}, {Li}, {Lian}, {Sheng}, {Li}, {Guo}, {Zhou}, {Zhang}, {Xu}, {Cheng}, \& {Liu}}]{2022Natur.610..656Z}
{Zhang}, L., {He}, J., {deBoer}, R.~J., {et~al.} 2022, \nat, 610, 656, \dodoi{10.1038/s41586-022-05230-x}

\bibitem[{{Zhen} {et~al.}(2023){Zhen}, {L{\"u}}, {Liu}, {Dohi}, {Nishimura}, {Zhu}, {Song}, {Wang}, \& {Xu}}]{2023ApJ...950..110Z}
{Zhen}, G., {L{\"u}}, G., {Liu}, H., {et~al.} 2023, \apj, 950, 110, \dodoi{10.3847/1538-4357/accd5f}

\bibitem[{{Zhou} {et~al.}(2023){Zhou}, {Wang}, {Zhang}, {Litvinov}, {Meisel}, {Blaum}, {Zhou}, {Hou}, {Li}, {Xu}, {Chen}, {Deng}, {Fu}, {Ge}, {He}, {Huang}, {Jiao}, {Li}, {Li}, {Liao}, {Litvinov}, {Liu}, {Niu}, {Shuai}, {Shi}, {Song}, {Sun}, {Wang}, {Xing}, {Xu}, {Xu}, {Yan}, {Yang}, {Yu}, {Yuan}, {Yuan}, {Zeng}, {Zhang}, \& {Zhang}}]{2023NatPh..19.1091Z}
{Zhou}, X., {Wang}, M., {Zhang}, Y.~H., {et~al.} 2023, Nature Physics, 19, 1091, \dodoi{10.1038/s41567-023-02034-2}

\end{thebibliography}
\bibliographystyle{aasjournal}

\end{document}